%% file: cepnf3clv.tex
\documentclass[%
superscriptaddress,
preprint,
preprintnumbers,
tightenlines,
showpacs,
showkeys,
nofootinbib,
amsmath,
amssymb,
aps,
prd,
floatfix,
]{revtex4-1}

\usepackage{graphicx}

\newcommand{\eval}[1] {\langle  #1 \rangle}

\newcommand{\tr}{{\rm Tr}}
\newcommand{\rmE}{{\rm E}}
\newcommand{\rmPS}{{\rm PS}}
\newcommand{\rmPSE}{{\rm PS, E}}
\newcommand{\rmt}{{\rm t}}
\newcommand{\rmPSt}{{\rm PS, t}}
\newcommand{\dof}{{\rm d.o.f.}}

\begin{document}
\title
{Critical endpoint of finite temperature phase transition for three flavor QCD} 

\author{Xiao-Yong Jin\footnote{Present address: Argonne Leadership Computing Facility, Argonne National Laboratory, Argonne, Illinois 60439, USA}}
\affiliation{RIKEN Advanced Institute for Computational Science, Kobe, Hyogo 650-0047, Japan}

\author{Yoshinobu Kuramashi}
\affiliation{Faculty of Pure and Applied Sciences, University of Tsukuba, Tsukuba, Ibaraki 305-8571, Japan}
\affiliation{Center for Computational Sciences, University of Tsukuba, Tsukuba, Ibaraki 305-8577, Japan}
\affiliation{RIKEN Advanced Institute for Computational Science, Kobe, Hyogo 650-0047, Japan}

\author{Yoshifumi~Nakamura}\email[]{nakamura@riken.jp}
\affiliation{RIKEN Advanced Institute for Computational Science, Kobe, Hyogo 650-0047, Japan}
\affiliation{Graduate School of System Informatics, Department of Computational Sciences, Kobe University, Kobe, Hyogo 657-8501, Japan}

\author{Shinji Takeda}
\affiliation{Institute of Physics, Kanazawa University, Kanazawa 920-1192, Japan}
\affiliation{RIKEN Advanced Institute for Computational Science, Kobe, Hyogo 650-0047, Japan}

\author{Akira Ukawa}
\affiliation{RIKEN Advanced Institute for Computational Science, Kobe, Hyogo 650-0047, Japan}

\date{\today}
\begin{abstract}
We investigate the critical endpoint of finite temperature phase transition of $N_f=3$ QCD at zero chemical potential.
We employ the renormalization-group improved Iwasaki gauge action and non-perturbatively $O(a)$-improved Wilson-clover  fermion action.
The critical endpoint is determined by using the intersection point 
of kurtosis for the temporal size $N_t$=4, 6, 8.  Spatial sizes of $N_l$=6--16 ($N_t$=4),  10--24 ($N_t$=6), and 12--24 ($N_t$=8) are employed.  We find that $N_t$=4 is out of the scaling region.   Using results for $N_t$=6 and 8, and making linear extrapolations in $1/N_t^2$, we obtain 
$\sqrt{t_0}T_\rmE=0.0975(14)(8)$, 
$\sqrt{t_0}m_\rmPSE=0.2254(52)(105)$ and 
$m_\rmPSE/T_\rmE=2.311(63)(13)$, 
where the first error is statistical error, the second error is systematic error, and $m_\rmPS$ is the pseudo scalar meson mass. 
 If one uses $1/\sqrt{t_0}=1.347(30)$~GeV reported by Borsanyi {\it et al.}, one finds
$T_\rmE=131(2)(1)(3)$~MeV, $m_\rmPSE=304(7)(14)(7)$~MeV and $m_\rmPSE/m_\rmPS^{\rm phys, sym}=0.739(17)(34)(17)$,
where the third error comes from error of $\sqrt{t_0}$ and $m_\rmPS^{\rm phys, sym}=\sqrt{(m_\pi^2+2m_K^2)/3}$.
Our current estimation of $\sqrt{t_0}m_\rmPSE$ in the continuum limit is about 25\% smaller than the SU(3) symmetric point.
\end{abstract}

\pacs{11.15.Ha,12.38.Gc,25.75.Nq}

\preprint{UTHEP-664, UTCCS-P-77, KANAZAWA-14-09}
\maketitle

\section{Introduction}\label{sec:int}

Knowledge of QCD phase structure is the basis for understanding the physics of the strong interactions at finite temperature and density.  Since simulations with finite chemical potential is plagued with the sign problem, it is important to clearly understand the finite temperature phase diagram as a function of light u-d quark masses and strange quark mass before starting extensive studies
at finite chemical potential.  

Analytical arguments indicate that the finite temperature transition with 3 massless quarks is of first order~\cite{PisarskiWilczek84}, which should then extend into the region of finite quark masses, ending at a line of critical points belonging to the Z(2) universality class~\cite{GavinGokschPisarski94}.  Theoretical considerations alone cannot tell if the physical point with $m_{ud}/m_s\approx 25$ lies on the crossover side or the first order side of the critical line. 

The results of simulations done to date to locate the critical line are rather confusing.   
All results with staggered fermion action are consistent with the physical point
being in the crossover region~\cite{Kaya99,Karsch01,Philipsen07,Smith11,Endrodi07,Ding11}.
However, on the location of the critical line, while the standard staggered action yields the critical quark mass as large as the physical up-down quark masses  both for 
$N_t$=4~\cite{Kaya99,Karsch01,Philipsen07,Smith11} and $N_t$=6~\cite{Philipsen07},  more recent studies with improved staggered actions indicate that the critical quark mass is significantly smaller than the physical up-down quark masses~\cite{Endrodi07,Ding11}. 
 In fact improved staggered actions have yet not found first order signals.  
On the other hand, with the Wilson quark action, an extensive pioneering study with naive Wilson action and $N_t$=4 temporal lattice size found that the physical point lies in the first order phase transition region~\cite{Iwasaki96}.

To clarify the issue, we study the critical endpoint for $N_f=3$ lattice QCD with non-perturbatively $O(a)$-improved Wilson fermion action, and attempt to estimate the continuum limit.    
Wilson type fermion action has exact flavor symmetry.  
In contrast, taste breaking with staggered type fermion action becomes large at coarse lattice spacings~\cite{HISQ2012}.  We consider this to be a potentially more serious issue in comparison to lack of chiral symmetry with Wilson type fermion actions.
Indeed the lack of chiral symmetry should not be a major problem for studying the critical endpoint which is expected to take place at a finite value of the quark mass. 

This paper is organized as follows.
In section~\ref{sec:method} we recapitulate finite size scaling of kurtosis which we employ to locate the critical endpoint. 
In section~\ref{sec:obs} we describe our observables, taking chiral condensate as an example.  
In section~\ref{sec:sim}  We  present the simulation details including the parameters and the simulation algorithm. 
Our numerical results are presented in section~\ref{sec:res}.  Our conclusions are summarized in section~\ref{sec:sum}.

\section{Method}
\label{sec:method}

Consider an observable ${\cal O}$.   If one denotes by $x$ the conjugate variable to ${\cal O}$, the $n$-th cumulant $\kappa_n$ of ${\cal O}$ is defined by
\begin{eqnarray}
\log\left<\exp\left(x{\cal O}\right)\right>=\sum_{n=1}^{\infty}\frac{x^n}{n!}\kappa_n.
\end{eqnarray}
The mean value $O$ of ${\cal O}$, susceptibility $\chi_{\cal O}$, skewness $S_{\cal O}$, and kurtosis $K_{\cal O}$ are defined as 
\begin{eqnarray}
O&=&\kappa_1/V=\left<{\cal O}\right>/V,\\
\chi_{\cal O}&=&\kappa_2/V=\left<\left({\cal O}-\left<{\cal O}\right>\right)^2\right>/V,\\
S_{\cal O}&=&\frac{\kappa_3}{\kappa_2^{3/2}},\\
K_{\cal O}&=&\frac{\kappa_4}{\kappa_2^2},
\end{eqnarray}
where $V$ is volume.
Of particular interest for locating the critical endpoint is kurtosis $K_{\cal O}$ since it is expected to have no or only small finite size effect at a second order transition point.  

Let $F(t,h,L)$ be the free energy of a system with a second order phase transition at $t=h=0$ and a linear extent $L$.  The two parameters $t$ and $h$ are reduced ``temperature'' and ``external field'' variables conjugate to the ``energy'' ${\cal E}$ and ``magnetization'' ${\cal M}$ operator characterizing the renormalization group flow around the critical point.  According to finite size scaling theory, the free energy scales as  $F(t,h,L)=F(tL^{y_t}, hL^{y_h}, 1)$ up to analytic terms.  If one approaches the critical point along the line $h=0$, the kurtosis for the magnetization ${\cal M}$ satisfies the scaling relation 
\begin{eqnarray}
K_{\cal M}=\frac{F^{(4)}(tL^{y_t})}{F^{(2)}(tL^{y_t})^2},
\end{eqnarray}
with $F^{(n)}(t)=\partial^nF(t,h,1)/\partial h^n\vert_{h=0}$.  Hence the kurtosis is expected to have a fixed value at $t=0$ independent of the size $L$.  

One can generalize the scaling relation to a more general class of operators ${\cal O}$ which may be expressed as a linear combination of ${\cal E}$ and ${\cal M}$ close to the second order transition point.  If the magnetization exponent $y_h$ is larger than the thermal exponent $y_t$, one finds 
\begin{eqnarray}
K_{\cal O}={\cal F}_{\cal O}(tL^{y_t})+L^{y_t-y_h} {\cal F'}_{\cal O}(tL^{y_t})+O(L^{2(y_t-y_h)}),
\end{eqnarray}
where ${\cal F}(x)$ is the scaling function for the leading term, and ${\cal F}'(x)$ that for the subleading term which is smaller by a factor $L^{y_t-y_h}$.  Thus there is a size-dependent  shift of kurtosis in the general case. 

At the first order phase transition point, for large volumes, kurtosis reaches the minimum~\cite{Billoire92} according to 
\begin{equation}
K_{\cal O} = -2 + {c \over N_l^d} + O(1/N_l^{2d})\,,
\end{equation}
where $d$ is the dimension.

\begin{figure}
\includegraphics[bb= 0 0 508 297,width=7.8cm]{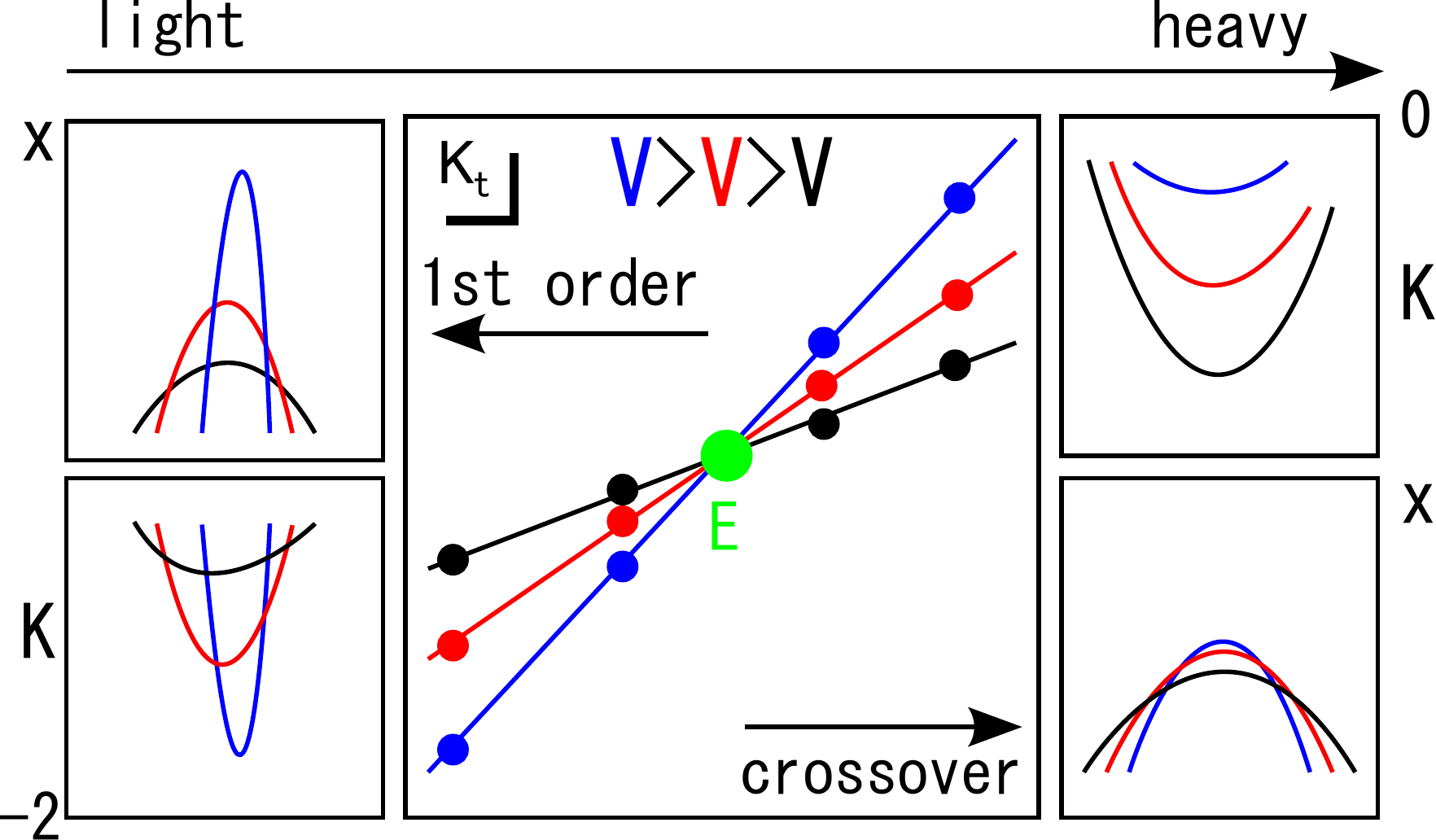}
\caption{\label{fig:method} Schematic picture of method to determine the critical endpoint.}
\end{figure}

We use the property of $K$ discussed above to determine the critical endpoint by the following strategy. 
\begin{enumerate}
\item For a given temporal lattice size $N_t$ and a spatial size $N_l$, we collect data for a set of  values of inverse gauge coupling $\beta$ and hopping parameter $\kappa$ around the critical endpoint.
\item For each $N_t$ and $N_l$, we determine the transition point  from the peak of susceptibility; this is done either by fixing $\beta$ and fitting the susceptibility as a function of $\kappa$, or by fixing $\kappa$ and searching the peak by reweighting in $\beta$~\cite{reweighting}. 
\item Using the value of $\kappa$ at the peak, we determine the kurtosis at the peak $K_\rmt$ for each spatial lattice size  $N_l$ as a function of $\beta$.
\item Plotting $K_\rmt$ as a function of $\beta$ for a set of spatial sizes $N_l$, we determine the critical endpoint by estimating the point where the kurtosis at different $N_l$ intersects; this is done by a fit inspired by the leading term of finite size scaling~\cite{Philipsen07}
\end{enumerate}
\begin{equation}\label{eq:fss_gfit}
K=K_\rmE + a N_l^{1/\nu}  (\beta-\beta_\rmE)\,,
\end{equation}
where $K_\rmE$ and $\beta_\rmE$ are the values of $K$ and $\beta$ at the critical endpoint, respectively.
In Fig.~\ref{fig:method} we sketch our method.

It is expected that the critical endpoint of $N_f=3$ (also $N_f=2+1$) QCD belongs to the universality class of the  3D Ising model with Z(2) symmetry~\cite{GavinGokschPisarski94}.
Therefore the values $K_\rmE=-1.396$ and $\nu=0.630$ are expected. However,  we do not assume any value for $K_\rmE$ and $\nu$ in the fitting.

\section{Observables}\label{sec:obs}

We can write the partition function with Wilson-clover fermion action as
\begin{equation}
Z   =\int dU    e^{-S_g} \prod_{i=1}^{N_f} \det D_i \,,
\end{equation}
where
\begin{equation}
\begin{split}
& D_i = {1\over 2\kappa_i } + C - {1\over 2} \sum_{\mu=1}^4 ( T^+_\mu + T^-_\mu) \,, \\
& C= \frac{i}{4}  \, c_{\rm SW}\, \sigma_{\mu\nu} F_{\mu\nu}(n) \delta_{m,n}\,, \\
& T^\pm_\mu = (1\mp \gamma_\mu) U_{\pm \mu}(n) \delta_{n,m\pm \hat{\mu}}\,, \\
& am_i  = {1 \over 2\kappa_i} + {\rm const.}
\end{split}
\end{equation}

Let us introduce the derivatives, 
\begin{equation}
\begin{split}
W_1(m_i)&={\partial   \log\det D_i(m_i) \over \partial  am_i   }=   \tr [   D^{-1}_i ] \,,\\
W_2(m_i)&={\partial   W_1(m_i) \over \partial  am_i   } = -  \tr [  D_i^{-2}  ] \,,\\
W_3(m_i)&={\partial^2 W_1(m_i) \over \partial (am_i)^2} = 2  \tr [  D_i^{-3}  ] \,,\\
W_4(m_i)&={\partial^3 W_1(m_i) \over \partial (am_i)^3} = -6 \tr [  D_i^{-4}  ] \,,\\
W_j(m_i)&={\partial^{j-1} W_1(m_i) \over \partial (am_i)^{j-1}}  \\
          &= (-1)^{j-1}  \times (j-1)! \times   \tr [  D_i^{-j} ] \,,\\
\end{split}
\end{equation}
and flavor averages, 
\begin{equation}
 \bar{W_j} = {1\over N_f} \sum_i^{N_f} W_j(m_i) \,.\\
\end{equation}
We consider derivatives of $X=\prod_i^{N_f} \det D_i $ with respect to $am_i$ up to 4th order.
Defining
\begin{equation}
\begin{split}
{ \partial^j \over \partial (am)^j } =  \Bigg({1\over N_f} \sum_i^{N_f} {\partial \over \partial am_i } \Bigg)^j \,,
\end{split}
\end{equation}
we find
\begin{equation}
\begin{split}
{ \partial   \over \partial (am)   } X & = X   \bar{W_1} \equiv XQ_1  \,,\\
{ \partial^2 \over \partial (am)^2 } X & = X  [\bar{W_2}  + \bar{W_1}^2 ] \equiv XQ_2  \,,\\
{ \partial^3 \over \partial (am)^3 } X & = X  [\bar{W_3}  + 3\bar{W_2} \bar{W_1} + \bar{W_1}^3 ]  \equiv XQ_3  \,,\\
{ \partial^4 \over \partial (am)^4 } X & = X  [\bar{W_4}  + 4\bar{W_3} \bar{W_1} +3\bar{W_2}^2 + 6\bar{W_2}\bar{W_1}^2 + \bar{W_1}^4 ] \equiv XQ_4  \,.\\
\end{split}
\end{equation}

We find then that the mean value $\Sigma$ and susceptibility $\chi_\Sigma$ are given by 
\begin{equation}
\begin{split}
& \Sigma    \equiv { 1 \over N_l^3 N_t } {\partial   \ln Z \over \partial  am    } = { \eval{Q_1} \over N_l^3 N_t }  \,, \\
& \chi_\Sigma \equiv { 1 \over N_l^3 N_t } {\partial^2 \ln Z \over \partial (am)^2 } = { \eval{Q_2} - \eval{Q_1}^2 \over N_l^3 N_t } \,, \\
\end{split}
\end{equation}
and  skewness $S_\Sigma$ and kurtosis $K_\Sigma$ by 
\begin{equation}
\begin{split}
S_\Sigma &= {\partial^3 \ln Z \over \partial (am)^3 } \Big/  \Big( {\partial^2 \ln Z \over \partial (am)^2 }  \Big)^{3 \over 2} 
               = {\eval{Q_3} -3 \eval{Q_2}\eval{Q_1} +2 \eval{Q_1}^3 \over (\eval{Q_2} - \eval{Q_1}^2)^{3/2}  } \,, \\
K_\Sigma &= {\partial^4 \ln Z \over \partial (am)^4 } \Big/  \Big( {\partial^2 \ln Z \over \partial (am)^2 }  \Big)^2 \\
  &= {\eval{Q_4} -4 \eval{Q_3}\eval{Q_1} -3\eval{Q_2}^2 +12\eval{Q_2}\eval{Q_1}^2  -6 \eval{Q_1}^4 \over (\eval{Q_2} - \eval{Q_1}^2)^2  } \,. \\
\end{split}
\end{equation}
For gluonic observables, gluon action density $G$, plaquette $P$ and Polyakov loop $L$, $Q_i$ is just the $i$-th power of the operator.

\section{Simulation Details}\label{sec:sim}

We choose three temporal lattice sizes $N_t$=4, 6, 8 to examine the continuum limit, 
and run simulations for a set of spatial sizes $N_l = 6-16$ ($N_t=4$), $10-24$ ($N_t=6$), and $12-24$ ($N_t=8$) for finite size scaling studies.
Calculations  are made with  $N_f=3$ degenerate flavors of dynamical quarks 
using the Iwasaki glue~\cite{iwasaki} and the nonperturbatively $O(a)$-improved Wilson fermion action~\cite{csw}, {\it i.e.},
we determine the critical endpoint on the line of $m_s = m_{ud}$ $(=m_l)$ on the Columbia phase diagram plot.

We use a highly optimized HMC code~\cite{BQCD}, applying mass preconditoning~\cite{mprec} and 
RHMC~\cite{RHMC}, 2nd order minimum norm integration scheme~\cite{Omelyan},
putting the pseudo fermion action on multiple time scales~\cite{mtime}
and a minimum residual chronological method~\cite{chronological} to choose the starting guess for the solver.

We generate O(100,000) trajectories for each lattice parameter set $(\beta, \kappa, N_t, N_l)$. 
We measure the gluon observables $G, P, L$ and cumulants at every trajectory, and the quark observable $\Sigma$ and cumulants at every 10 trajectories.
Errors of the observables are estimated by jackknife method with the bin size of $O(1,000)$ trajectories.   

\begin{figure}[b]
\includegraphics[bb= 0 0 454 340,width=6.5cm]{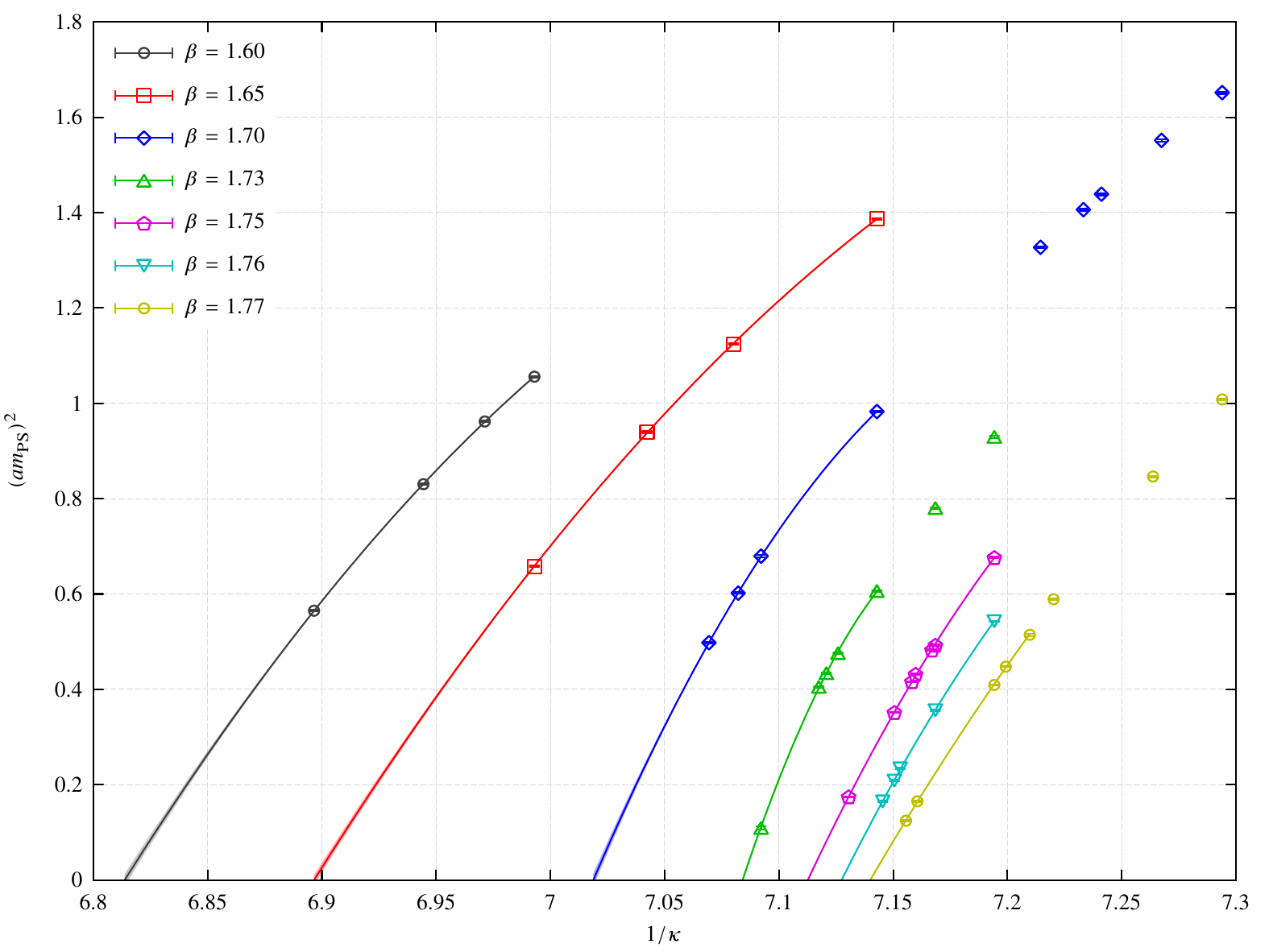}
\includegraphics[bb= 0 0 454 340,width=6.5cm]{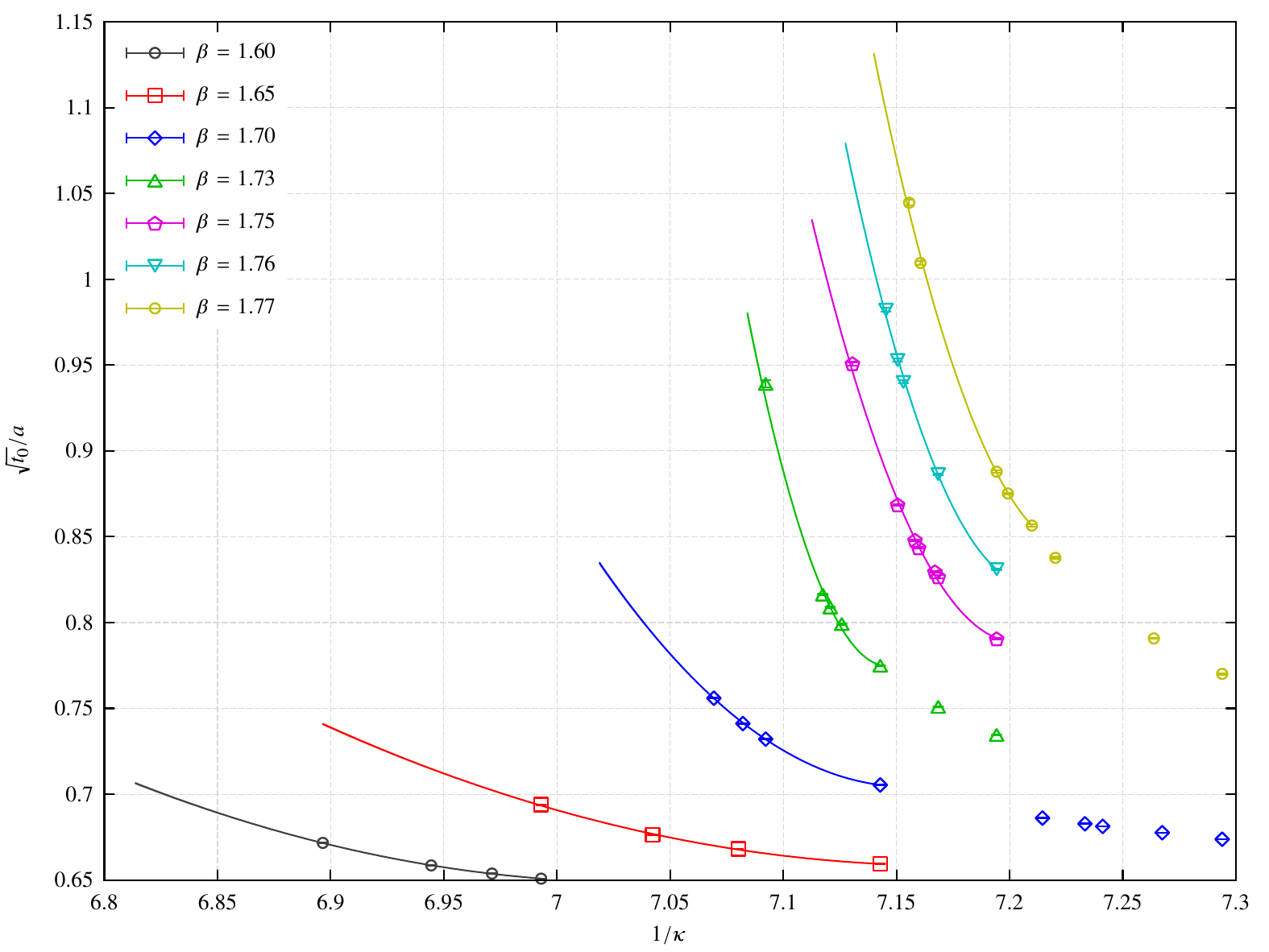}
\caption{\label{fig:scale} Left: $(am_{\rm PS})^2$  v.s $1/\kappa$. Right: $\sqrt{t_0}/a$  v.s $1/\kappa$.}
\end{figure}

To set the physical scale, we perform a set of zero temperature simulations.
The simulation parameters, results for mass of pseudo-scalar meson and the Wilson flow scale parameter $\sqrt{t_0}/a$~\cite{wilflow} are summarized in Tables~\ref{tab:scale160} -- ~\ref{tab:scale177} in Appendix~\ref{sec:scale}.
Data are  based on $O(500)$ configurations separated by 10 trajectories.  Errors are estimated by the jack knife method with a bin size of 10 to 100 configurations.   

Using a combined fit of form,
\begin{eqnarray}
\label{eq:scalefit-1}
 (am_{\rm PS})^2 &=& a_1 ({1\over \kappa} - {1 \over \kappa_c}) + a_2 ({1\over \kappa} - {1\over \kappa_c})^2 \,, \\
 \label{eq:scalefit-2}
{\sqrt{t_0} \over a} &=& b_0 + b_1 ({1\over \kappa} - {1 \over \kappa_c}) + b_2 ({1\over \kappa} - {1 \over \kappa_c})^2\,,
\end{eqnarray}
we obtain the critical hopping parameter $\kappa_c$ and the expansion coefficients. 
The fit results are listed in Table~\ref{tab:scalefit} in Appendix~\ref{sec:scale}.   We plot data and fits for the pseudo scalar meson mass and Wilson flow scale in Fig.~\ref{fig:scale}. 

Our calculations are carried out on the K computer provided by the RIKEN Advanced Institute for Computational Science,
HA8000 and FX10 at University Tokyo, HA8000-tc/HT210 and FX10 at Kyushu University and 
Cray XC30 with Xeon Phi at Kyoto University.

\section{Results}\label{sec:res}

\subsection{Location of finite temperature transition}

The first step of analysis is to locate the position of thermal transition on the $(\beta, \kappa)$ plane for each $N_t$ and $N_l$.  In Appendix \ref{sec:observablefigure} we show representative results of susceptibility and kurtosis for the four observables at $N_t=4$ (Figs.~\ref{fig:nt4_b160} and \ref{fig:nt4_b165}), $N_t=6$ (Figs.~\ref{fig:nt6_b1715} and \ref{fig:nt6_b173}), and $N_t=8$ (Figs.~\ref{fig:nt8_k014024} and \ref{fig:nt8_k013995}).  The pairs of figures are chosen so that the pair sandwiches the critical endpoint. 

For $N_t=4$ and 6, we fix $\beta$ and make measurements at several values of $\kappa$.  In order to locate the transition point, susceptibility and kurtosis are fitted with a quadratic ansatz in $\kappa$.  As one observes from the fit curves, the maximum of susceptibility and the minimum of kurtosis are mutually consistent. Though not shown, we find that the zero of skewness obtained by a linear fit is  consistent as well.   We choose the maximum of susceptibility to be the point of transition.  The results for the value of $\kappa$ thus defined is listed in Table~\ref{tab:kappat46} in Appendix~\ref{sec:transition}.  The errors are calculated from the fit.  

For $N_t=8$, we fix $\kappa$ and make runs at several values of $\beta$.  Since these runs are computationally expensive, we carry out long runs at one or two judiciously chosen values of $\beta$ close to the would-be transition point and make single- or multi-ensemble reweighting to search for the maximum susceptibility as shown in Figs.~\ref{fig:nt8_k014024} and \ref{fig:nt8_k013995} in Appendix~\ref{sec:observablefigure}.   The values of $\beta$ corresponding to the susceptibility peak are listed in Table~\ref{tab:betat8}.  The central values and errors are calculated by estimating the errors of reweighted values by jackknife method with bin size of 100 trajectories.  

Having determined the location of thermal transition from the peak of susceptibility, we calculate the value of kurtosis at the transition point with the help of the fit or reweighting of kurtosis.  The results are listed in Tables~\ref{tab:krt46} and \ref{tab:krt8} in Appendix~\ref{sec:transition}. 

\subsection{Kurtosis intersection analysis}

\begin{figure}
\includegraphics[bb= 0 0 340 255,width=5.8cm]{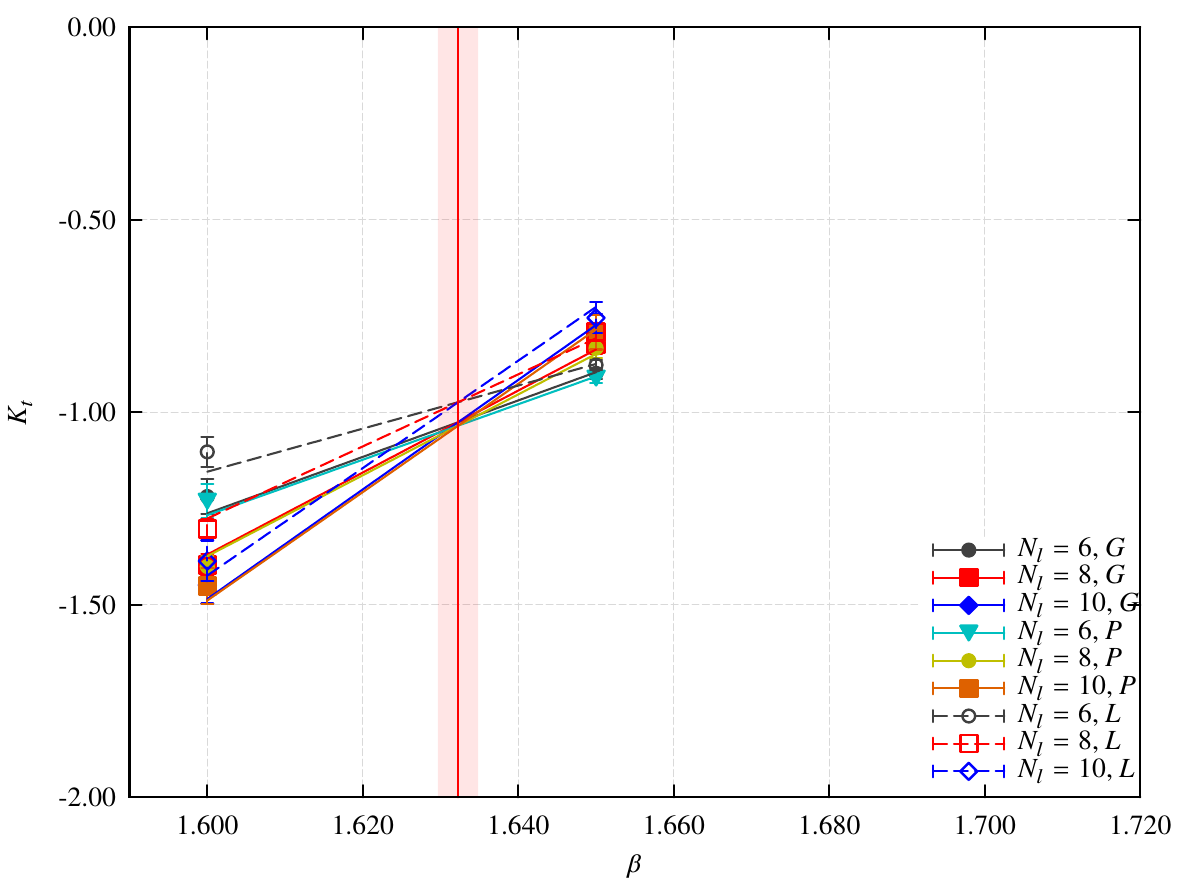}
\includegraphics[bb= 0 0 340 255,width=5.8cm]{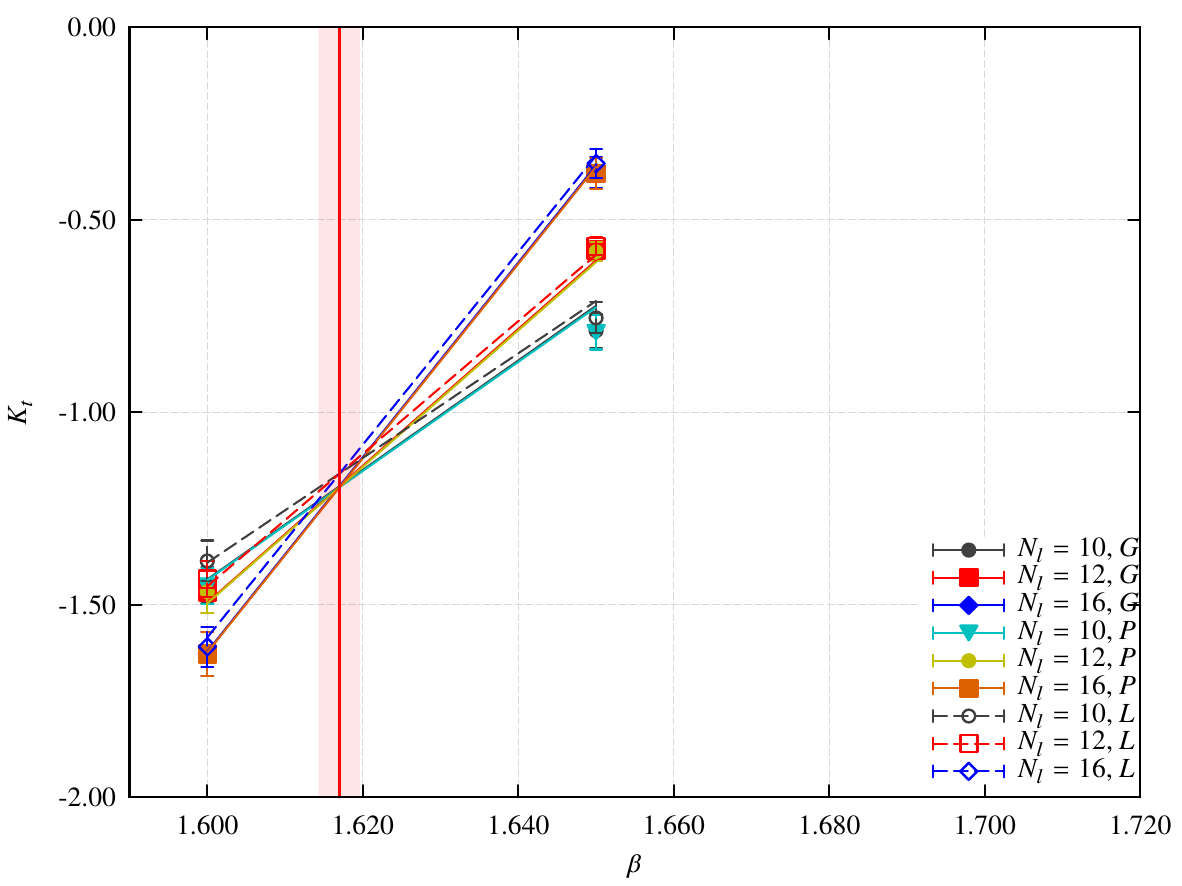}\\
\includegraphics[bb= 0 0 340 255,width=5.8cm]{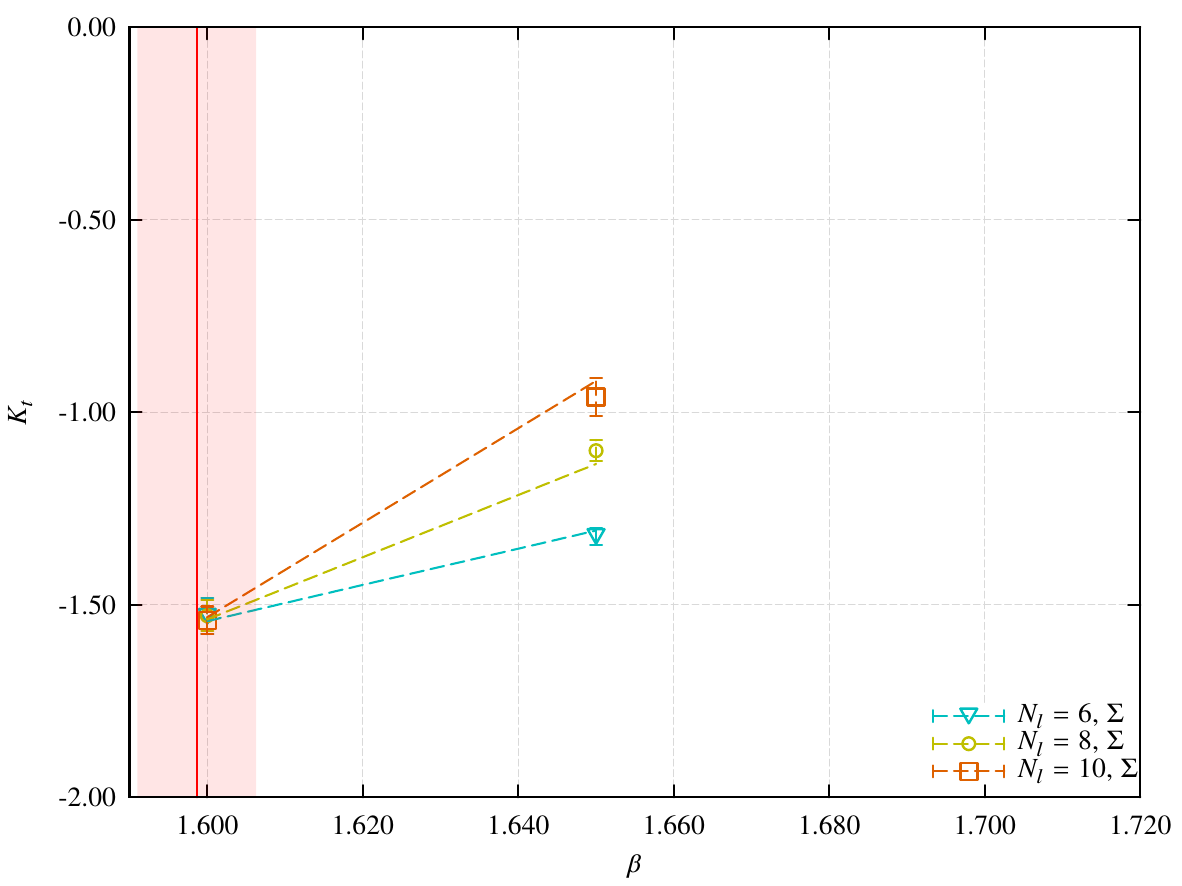}
\includegraphics[bb= 0 0 340 255,width=5.8cm]{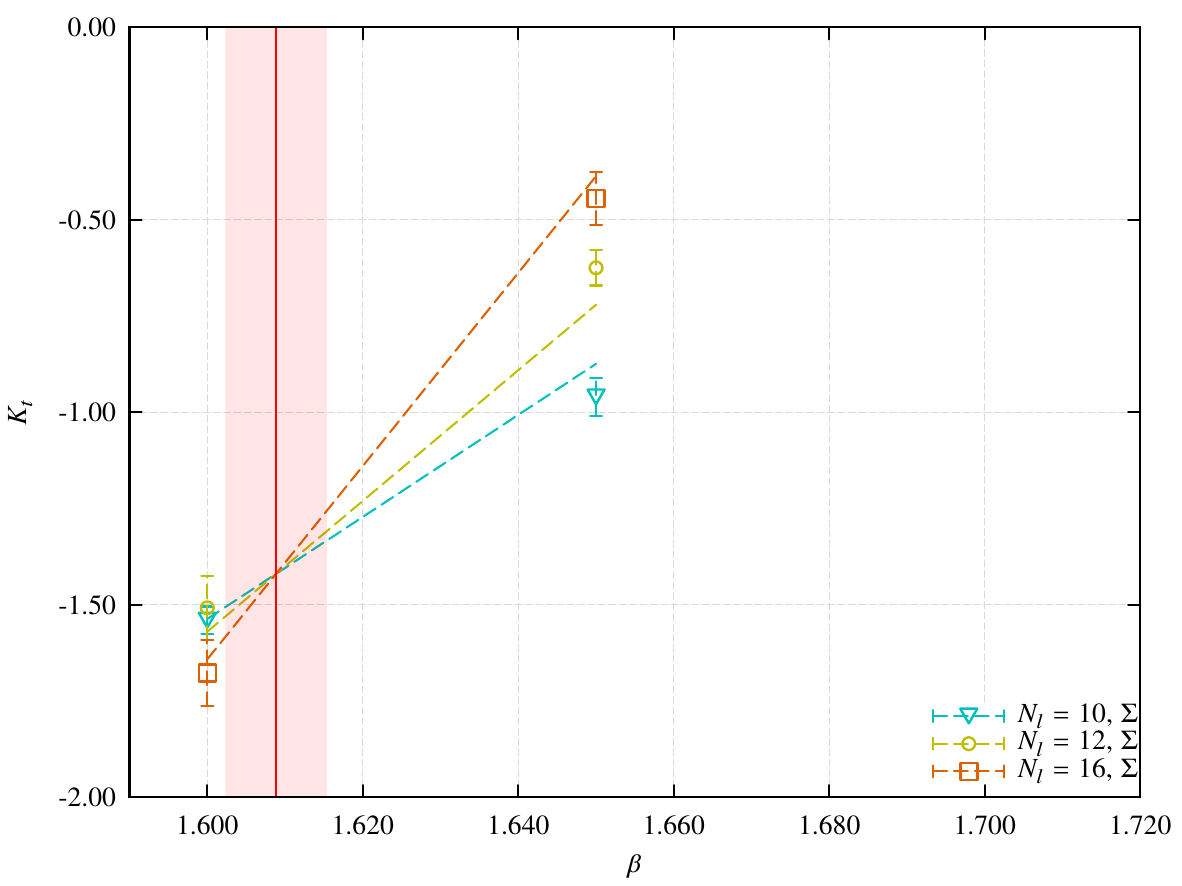}
\caption{\label{fig:cep_nt4vol}  Kurtosis intersection for $G, P, L$ on smaller lattices (top-left), larger lattices (top-right),
$\Sigma$ on smaller lattices (bottom-left), larger lattices (bottom-right) at $N_t=4$.}
\end{figure}

Ideally, if one plots the value of kurtosis along the transition line  with a given temporal size $N_t$, the curves for various spatial sizes $N_l$ intersects at a single point corresponding to the critical endpoint.  In practice, the intersection point varies for various combinations of $N_l$'s, and also  depends on the observables. 
This is illustrated in Fig.~\ref{fig:cep_nt4vol} where we separately plot $N_t=4$ results for the gluon observables $G, P, L$ for smaller spatial sizes $N_l=6, 8, 10$ and for larger sizes $N_l=10, 12, 16$, and similarly for the quark observable $\Sigma$.  We find that gluon observable and quark observable
give rather different estimates for the intersection point  $\beta_\rmE$ at smaller values of $N_l$'s which, however, becomes consistent for larger values of $N_l$'s.  

\begin{figure}
\includegraphics[bb= 0 0 340 255,width=7cm]{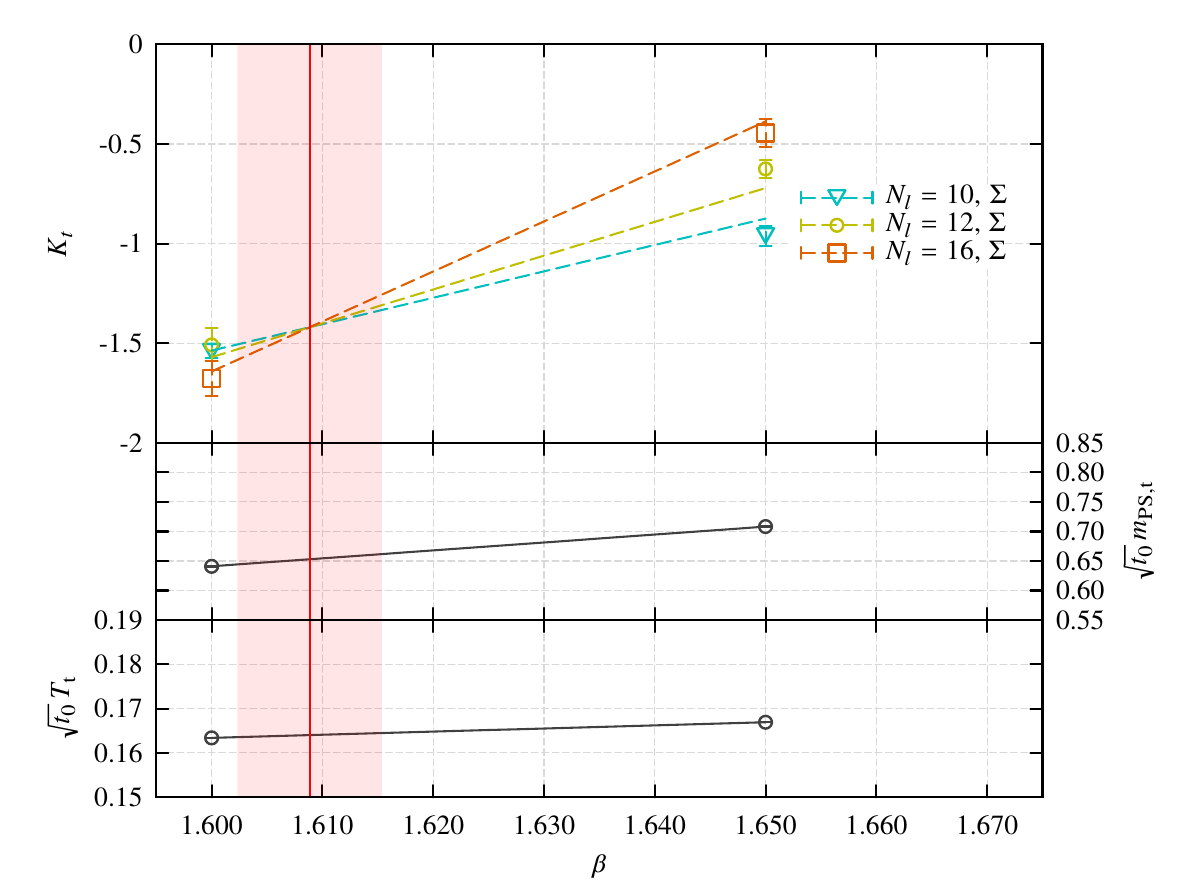}
\includegraphics[bb= 0 0 340 255,width=7cm]{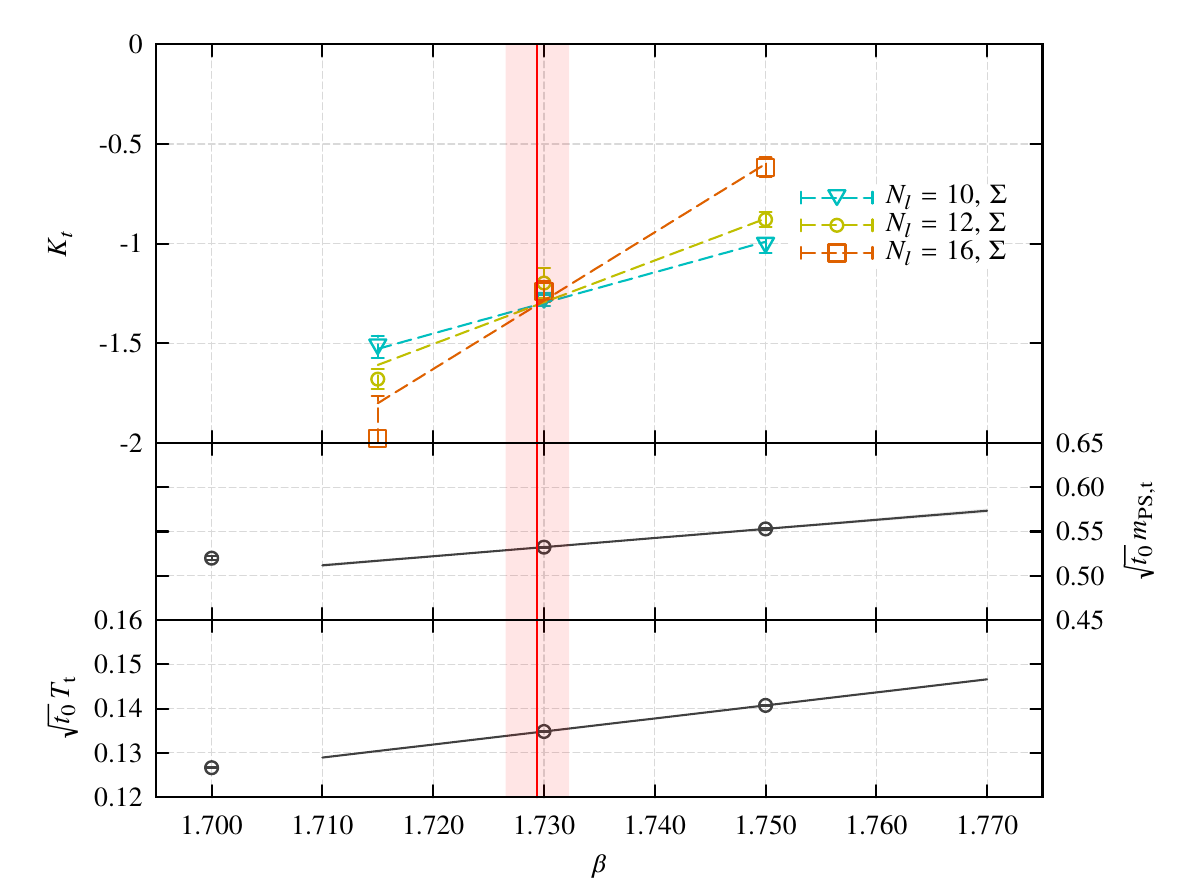}
\includegraphics[bb= 0 0 340 255,width=7cm]{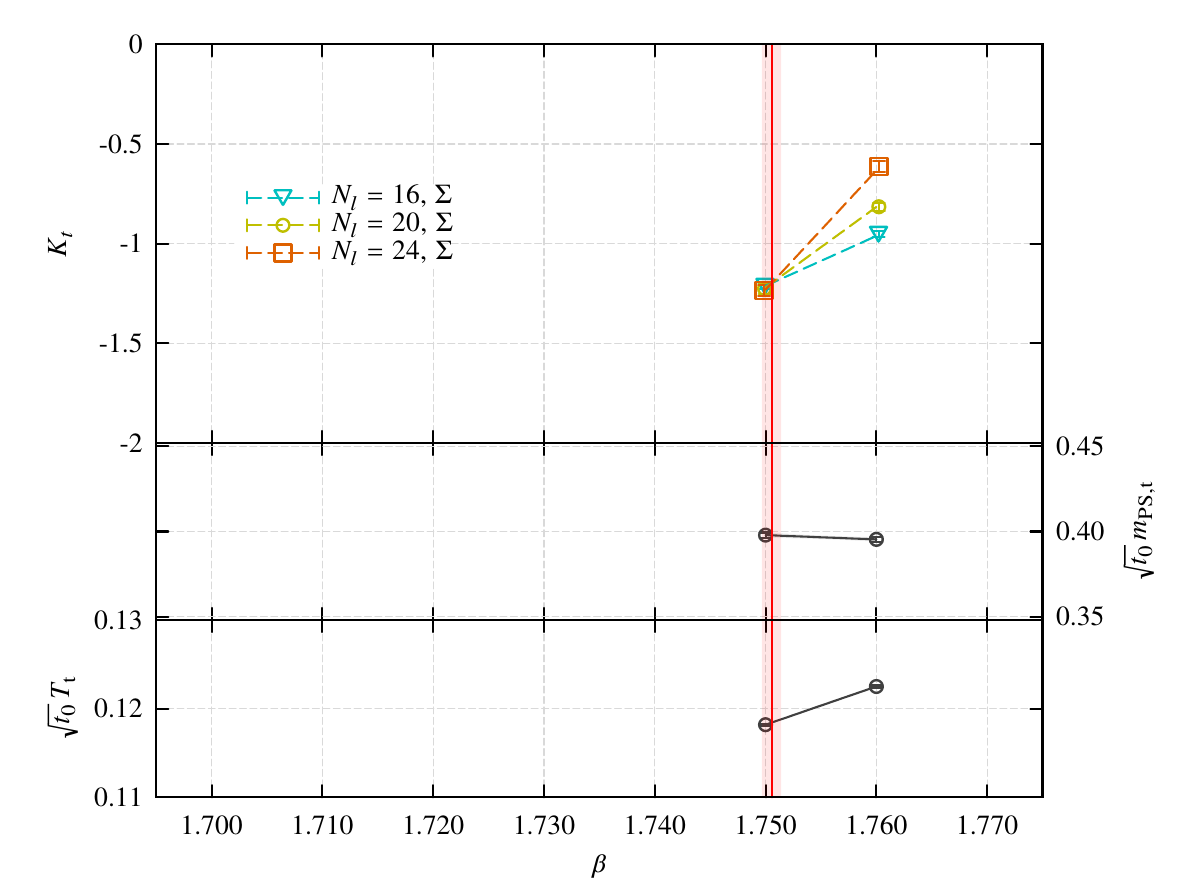}
\caption{\label{fig:cep_nt468} $K_\rmt$, $\sqrt{t_0} m_\rmPSt$, $\sqrt{t_0}T_\rmt$ v.s. $\beta$ at $N_t=4$(top-left), $N_t=6$(top-right), 
$N_t=8$(bottom) determined by the quark observable $\Sigma$.}
%
\includegraphics[bb= 0 0 340 255,width=7cm]{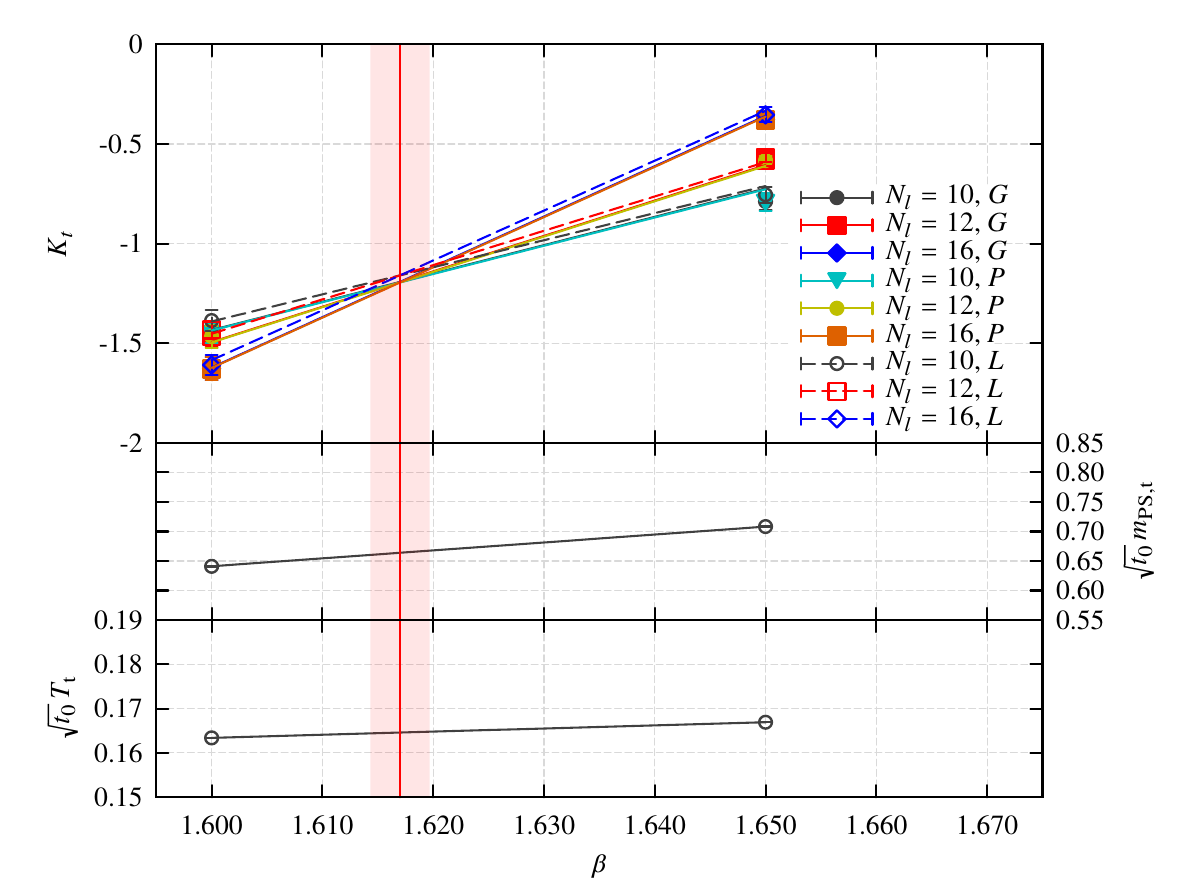}
\includegraphics[bb= 0 0 340 255,width=7cm]{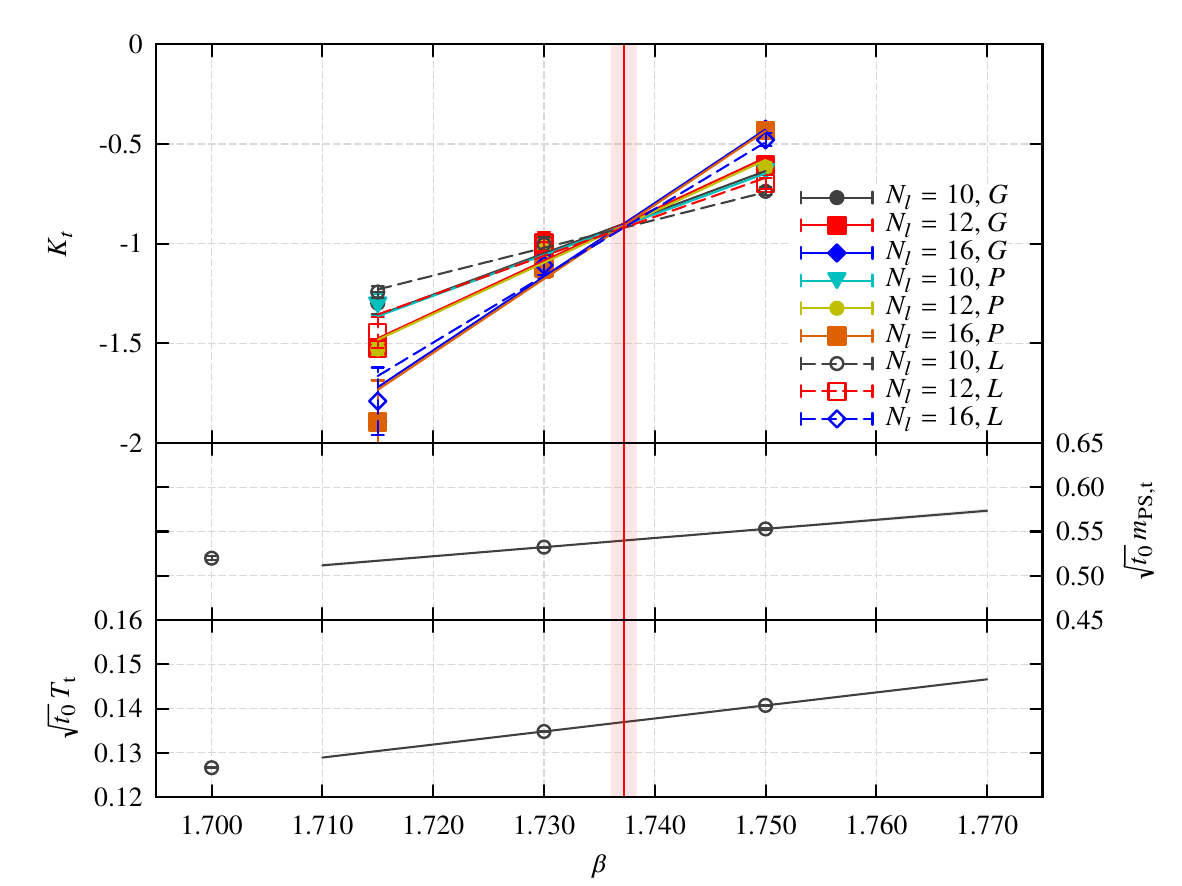}
\includegraphics[bb= 0 0 340 255,width=7cm]{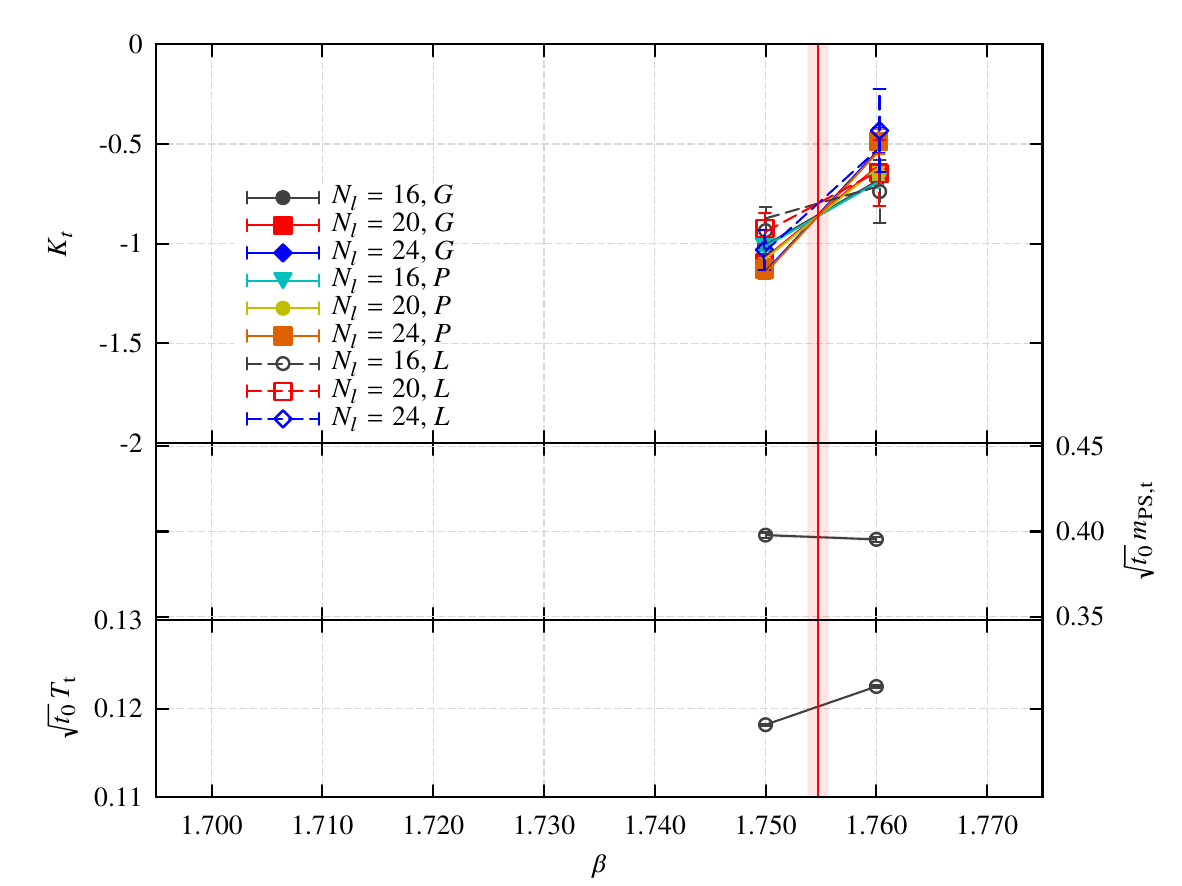}
\caption{\label{fig:cep_gluon_nt468} $K_\rmt$, $\sqrt{t_0} m_\rmPSt$, $\sqrt{t_0}T_\rmt$ v.s. $\beta$ at $N_t=4$(top-left), $N_t=6$(top-right), 
$N_t=8$(bottom) determined by the gluon observables $G, P, L$.}
\end{figure}

We suspect that the situation above is due to the fact that our observables are linear combinations of ``energy'' and ``magnetization'' operators with different mixing coefficients so that the sub leading corrections of $O(L^{y_t-y_h})$ differ in magnitude from observable to observable, especially between gluon and quark observables.  Since we do not have enough data set to resolve the sub leading contributions, we use the quark observable $\Sigma$ as our primary operator and employ gluon observables to estimate the systematic errors in the results. 

In Fig.~\ref{fig:cep_nt468} we show the intersection analysis using $\Sigma$.  Also plotted is the determination of the pseudo scalar meson mass $m_{\rm PS, E}$ and the transition temperature $T_{\rm E}$ at the critical endpoint  in units of the Wilson flow scale $\sqrt{t_0}$.   Figures~\ref{fig:cep_gluon_nt468} show similar plots for gluon observables.  The numerical values of the estimated critical endpoint $(\beta_E, \kappa_E)$, kurtosis at that point $K_{\rm E}$, the exponent $\nu$, the pseudo scalar meson mass $\sqrt{t_0}m_{\rm PS, E}$ and critical temperature $\sqrt{t_0}T_{\rm E}$ are listed in Table~\ref{tab:gfit}. 

\begin{table}
\caption{\label{tab:gfit}  Fit results for kurtosis intersection point and interpolated values of $\sqrt{t_0} m_\rmPS$ and $\sqrt{t_0} T$.  Top table shows values obtained with quark observable $\Sigma$.  Bottom table is obtained with gluon observables $G, P, L$.  We also list the value of $\kappa_E$ for completeness.}
\begin{ruledtabular}
\begin{tabular}{c|ccccc|c||c|c}
\multicolumn{8}{l}{Analysis with quark observable $\Sigma$}\\
\colrule
$N_t$  &$\beta_\rmE$ & $\kappa_E$ & $K_{\rmE}$ & $\nu$  & $a$  & $\chi^2/\dof$ & $ \sqrt{t_0} m_\rmPSE$ & $\sqrt{t_0}T_\rmE$ \\
\colrule
 4 &      1.6089(66) &     0.14305(30) &  -1.42(11) &   0.74(15) &   0.58(39) & 4.401 &      0.6530(88) &     0.16402(47) \\
 6 &      1.7294(29) &     0.14045(11) & -1.307(57) &   0.58(11) &   0.30(24) & 1.145 &      0.5317(31) &     0.13467(85) \\
 8 &     1.75054(87) &    0.140226(25) & -1.198(33) &  0.465(58) &  0.065(53) & 0.148 &      0.3977(16) &     0.11842(39) \\
%
\colrule
\colrule
\multicolumn{8}{l}{Analysis with gluon observables $G, P, L$}\\
\colrule
$N_t$  &$\beta_\rmE$ & $\kappa_E$ & $K_{\rmE}$ & $\nu$  & $a$  & $\chi^2/\dof$ & $ \sqrt{t_0} m_\rmPSE$ & $\sqrt{t_0}T_\rmE$ \\
\colrule
$$ &              $$ &              $$ & -1.192(54) &   0.82(13) &   0.85(41) &    $$ &              $$ &              $$ \\
 4 &      1.6170(27) &     0.14269(12) & -1.195(54) &   0.81(13) &   0.84(40) & 1.313 &      0.6639(37) &     0.16460(19) \\
$$ &              $$ &              $$ & -1.160(54) &   0.77(11) &   0.68(33) &    $$ &              $$ &              $$ \\
\colrule
$$ &              $$ &              $$ & -0.900(31) &   0.81(13) &   1.20(56) &    $$ &              $$ &              $$ \\
 6 &      1.7372(12) &    0.140163(44) & -0.910(31) &   0.79(12) &   1.11(52) & 1.469 &      0.5397(14) &     0.13696(35) \\
$$ &              $$ &              $$ & -0.920(25) &  0.536(63) &   0.19(11) &    $$ &              $$ &              $$ \\
\colrule
$$ &              $$ &              $$ & -0.856(40) &   0.65(17) &   0.43(52) &    $$ &              $$ &              $$ \\
 8 &     1.75474(96) &    0.140106(28) & -0.863(40) &   0.64(17) &   0.43(51) & 0.684 &      0.3967(11) &     0.12023(42) \\
$$ &              $$ &              $$ & -0.799(63) &   0.35(30) &  0.006(43) &    $$ &              $$ &              $$ \\
\end{tabular}
\end{ruledtabular}

\end{table}

We observe in this table that the kurtosis $K_{\rm E}$ and the exponent $\nu$ show significant variation depending on the choice of observables, particularly with $K_{\rm E}$.  This of course should not be so, and we likely need larger spatial sizes  to make certain that infinite volume values are attained. 
Comparing with the values of possible universality classes, {\it i.e.,} 
$K_\rmE=-1.396$ and $\nu=0.630$ for Z(2), 
$K_\rmE=-1.758$ and $\nu=0.672$ for O(2), and  
$K_\rmE=-1.908$ and $\nu=0.748$ for O(4), our numbers favor Z(2).  

\begin{figure}[t]
\includegraphics[bb= 0 0 454 340,width=8cm]{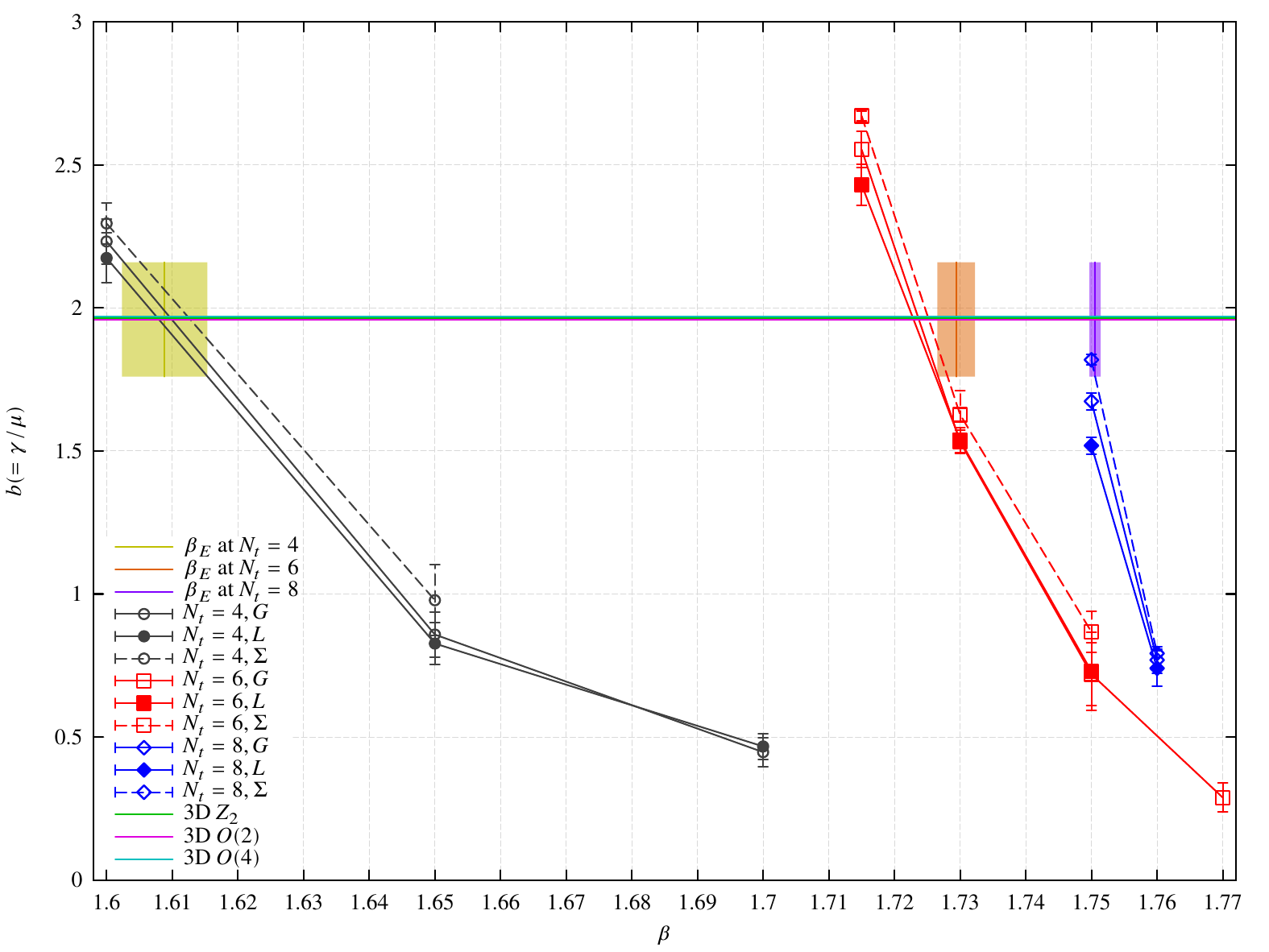}
\caption{\label{fig:exp} Susceptibility exponent $b$ as a function of $\beta$.  Horizontal line corresponds to the value $\gamma/\nu=1.964$ for Z(2) universality in 3 dimensions.  Colored blocks superimposed are estimates of the critical endpoint $\beta_E$ obtained in the kurtosis intersection analysis. }
\end{figure}

We may attempt to check the location of the critical endpoint and the universality class by looking at the spatial size variation of the peak height of susceptibility:
\begin{equation}
 \chi^{\rm max}=aN_l^b. 
\end{equation}
The exponent $b$ should equal $\gamma/\nu$ at the critical endpoint, $b=d$ the space dimension at the first order side at lower $\beta$, and $b=0$ at larger $\beta$ after the phase transition terminates.   
We plot the exponent $b$ as a function of  $\beta$ for each $N_t$ in Fig.~\ref{fig:exp}.   The horizontal line corresponds to the value $\gamma/\nu=1.964$ for Z(2).  Superimposed blocks on the line are the estimate  of $\beta_E$ from kurtosis intersection analysis in Table~\ref{tab:gfit}. 
We observe consistency with the crossing point of the exponent $b$ with the horizontal line.  Unfortunately, $\gamma/\nu$ has a similar value $\gamma/\nu=1.964, 1.962, 1.975$ for Z(2),  O(2) and O(4).  Distinguishing Z(2) from the other universality classes would be difficult. 

\begin{figure}
\includegraphics[bb= 0 0 340 255,width=8cm]{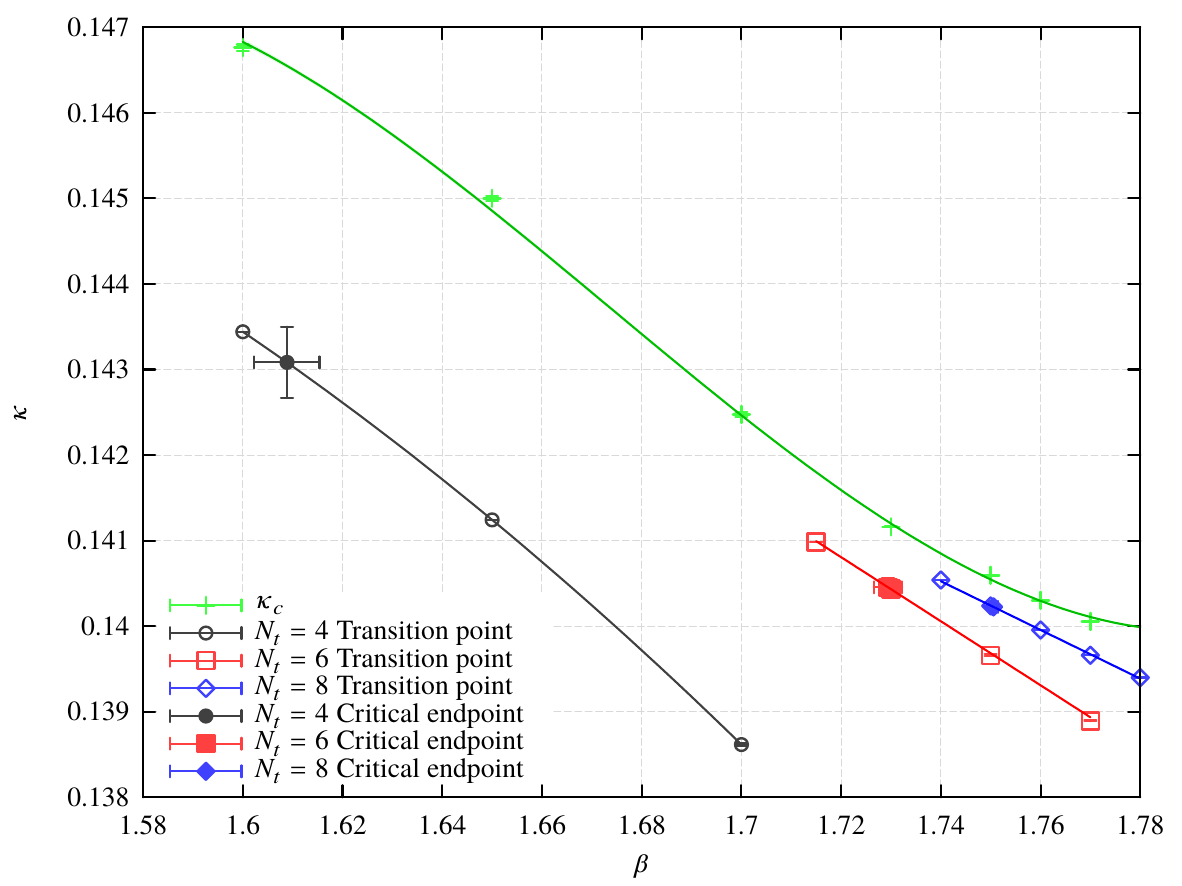}
\caption{\label{fig:bk} Phase diagram of $N_f=3$ lattice QCD on the $(\beta, \kappa)$ plane. Estimation of the line of first order phase transition ending at a critical point marked by full circles are shown for $N_t=4,6,8$, together with the line of $\kappa_c$ where pion mass vanishes. }
\end{figure}

In Fig.~\ref{fig:bk}, we show the phase diagram in the $(\beta, \kappa)$ plane.  Open symbols are the location of thermal transition as determined by the susceptibility peak analysis, and filled symbols are the location of the critical endpoint estimated by the kurtosis intersection analysis for $N_t=4, 6, 8$.  Lines are smooth interpolation of the thermal phase transition line; the segment to the left of the filled symbols corresponds first order transition and to the right is crossover. The location of critical hopping parameter $\kappa_c$ where pion mass vanishes is also shown.   

\subsection{Continuum estimation of  $m_\rmPS$ and  $T$ at the critical endpoint}

\begin{figure}
\includegraphics[bb= 0 0 340 255,width=7cm]{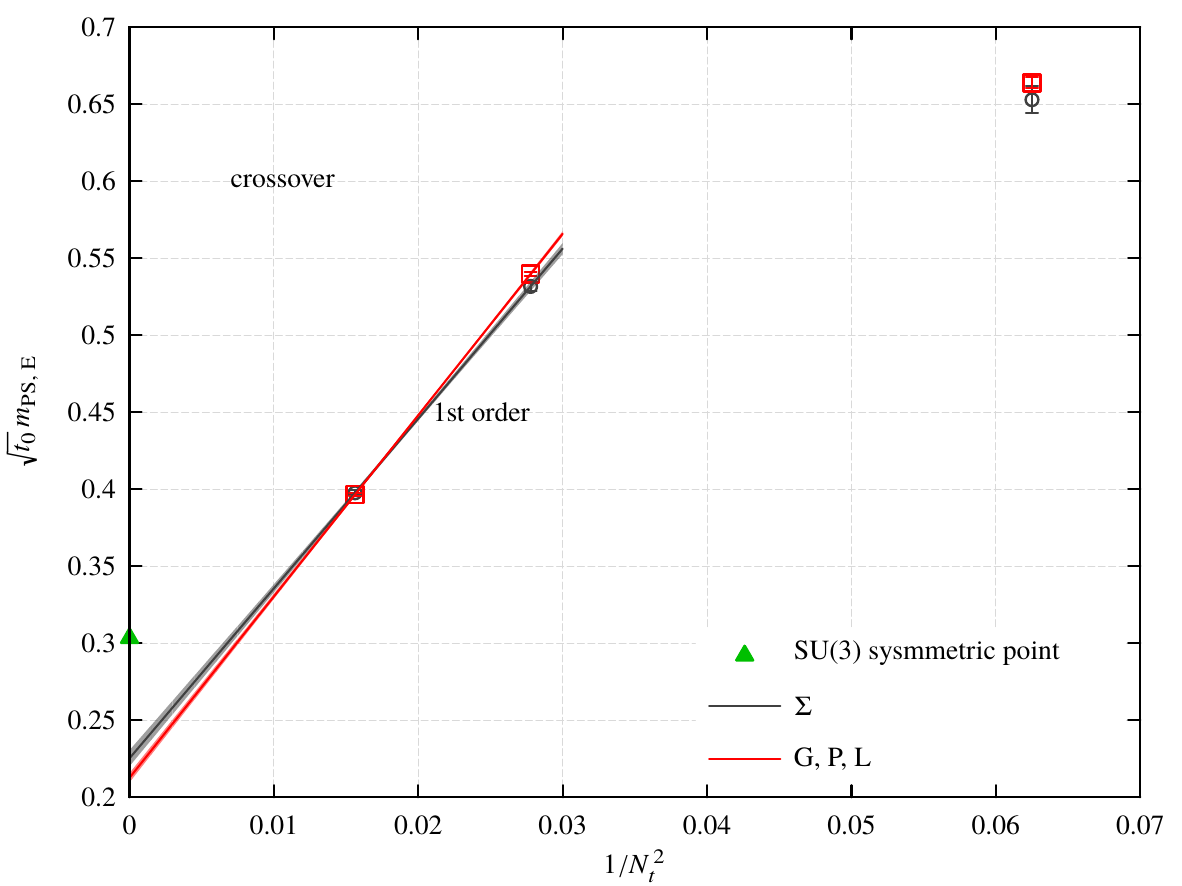}
\includegraphics[bb= 0 0 340 255,width=7cm]{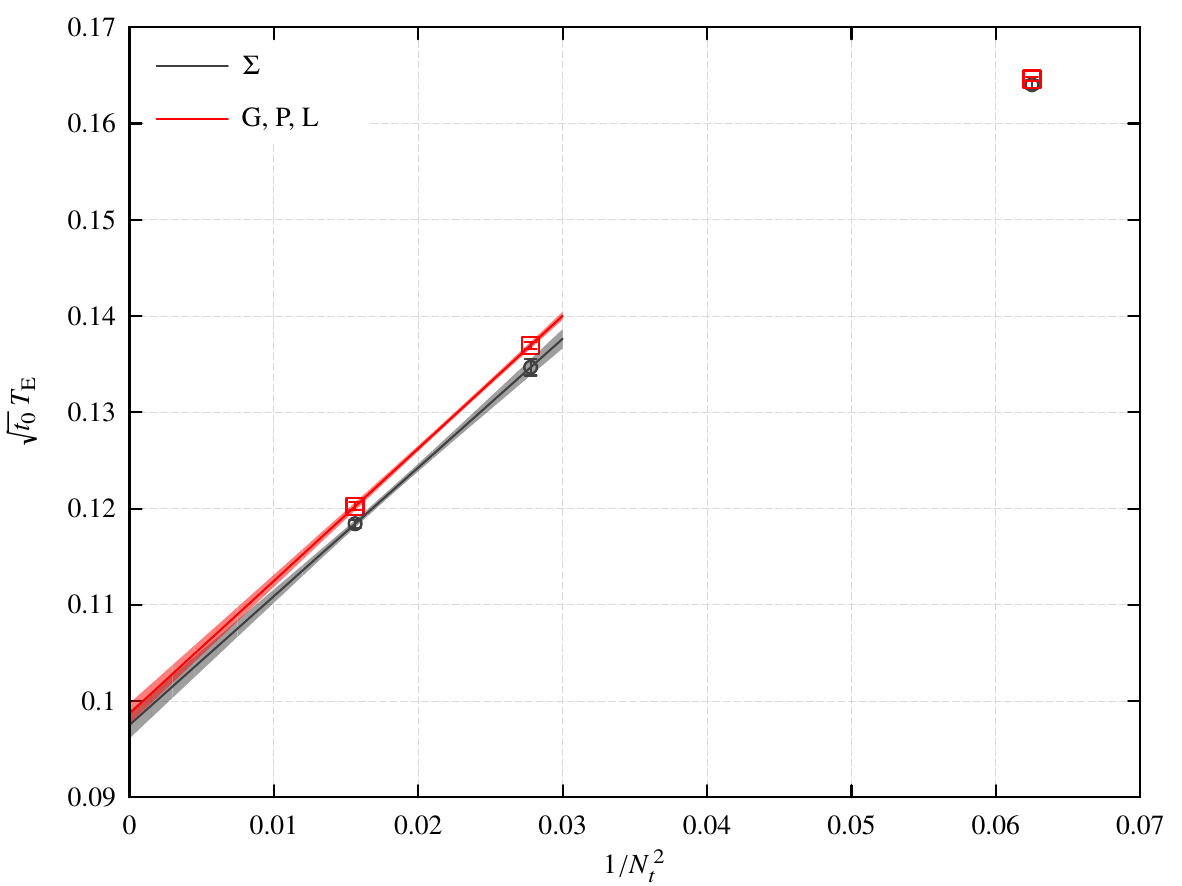}
\caption{\label{fig:continuum_limit} Continuum extrapolation of $\sqrt{t_0} m_\rmPSE$ and  $\sqrt{t_0}\,T_\rmE$ as a function of $1/N_t^2$. }
\end{figure}

Finally, we extrapolate  $\sqrt{t_0} m_\rmPSE$ and $\sqrt{t_0} T_\rmE$ to the continuum limit.
In Fig.~\ref{fig:continuum_limit} we plot $\sqrt{t_0} m_\rmPSE$ and $\sqrt{t_0} T_\rmE$ against $1/N_t^2$.  Results based on quark and gluon estimation of $\beta_E$ are both shown. 
The green triangle on the y-axis denotes the SU(3) symmetric point estimated by $\sqrt{t_0}\sqrt{(m_\pi^2 + 2m_K^2)/3}\sim 0.305$. 
We have used the value $\sqrt{t_0}=1.347(30)$~GeV~\cite{t0BMW} (see also \cite{t0QCDSF} ). 

It is clear that $N_t=4$ is outside the scaling region.  We therefore make a linear extrapolation with the points for $N_t=6$ and 8 to obtain
\begin{eqnarray}
\sqrt{t_0} m_{\rm PS, E}&=& 0.2254(52)(105) \,,\\
\sqrt{t_0} T_{\rm E} &=& 0.0975(14)(8) \,,
\end{eqnarray}
where the first error is statistical error and the second error is systematic error estimated from the spread between the determinations with quark and gluon observables. 
Normalizing by $m_\rmPS^{\rm phys, sym} \equiv \sqrt{(m_\pi^2 + 2m_K^2)/3}$, we obtain
\begin{equation}
{m_\rmPSE \over m_\rmPS^{\rm phys, sym}} = 0.738(17)(34)(17),
\end{equation}
where the third error comes from error of $\sqrt{t_0}$.
Converting to physical units, we estimate $T_\rmE=131(2)(1)(3)$~MeV and $m_\rmPSE=304(7)(14)(7)$~MeV. 

\section{Summary}\label{sec:sum}

We have investigated the critical endpoint of QCD with $N_f=3$ dynamical flavors of
non-perturbatively $O(a)$-improved Wilson fermions.
We have determined the critical endpoint by using the intersection
points of kurtosis at the temporal sizes $N_t=4, 6, 8$. 

The values of kurtosis $K_E$ at the critical endpoint and the exponent $\nu$ obtained from the intersection analysis favor Z(2) universality class,  and the location of the critical endpoint is consistent with those estimated from the susceptibility analysis.  Further work with larger spatial lattice sizes would be needed, however,  to establish the universality class. 

Our current estimate of the pseudo scalar meson mass at the critical endpoint in the continuum limit is about 25\% lighter than the flavor SU(3) symmetric point at the quark mass $m_q = (m^{phy}_u + m^{phy}_d + m^{phy}_s)/3$.
This suggests the possibility that the quark mass at the critical endpoint is not so small as suggested by recent studies with improved staggered quark action. 

Our continuum extrapolation has been made, however, only with two temporal lattice sizes $N_t=6$ and 8 as we have found that $N_t=4$ is out of the region of $1/N_t^2$ linear scaling.
We are planning further studies at larger temporal sizes to obtain conclusive results. 
 
\begin{acknowledgments}
The BQCD code~\cite{BQCD} was used in this work.
This research used computational resources of the K computer provided by the RIKEN Advanced Institute for Computational Science
through the HPCI System Research project (Project ID:hp120115), FX10 at University Tokyo and Kyushu University,
HA8000 at University of Tokyo through the HPCI System Research project (Project ID:hp130092) and System E at 
Kyoto University through the HPCI System Research project (Project ID:hp140180).
This work was supported by the Large Scale Simulation Program No.12/13-13(FY2012-2013) and No.13/14-12(FY2013-2014) of High Energy Accelerator Research Organization (KEK).
A part of this work is supported by the Grants-in-Aid for
Scientific Research from the Ministry of Education, 
Culture, Sports, Science and Technology 
(Nos. 26800130).

\end{acknowledgments}

\clearpage
\appendix

\section{Wilson flow scale and pseudo scalar meson mass at zero temperature}
\label{sec:scale}

Simulation parameters, results for mass of pseudo-scalar meson $am_{\rm PS}$, and Wilson flow scale parameter $\sqrt{t_0}/a$~\cite{wilflow} are summarized in Tables~\ref{tab:scale160} -- ~\ref{tab:scale177}. 
Result of fitting data to (\ref{eq:scalefit-1}) and (\ref{eq:scalefit-2})  as a function of $\kappa$ for each $\beta$ is given in Table~\ref{tab:scalefit}

\input{scale_table}

\clearpage
\section{Transition point and kurtosis}
\label{sec:transition}

Estimated transition point and the value of kurtosis at transition point $K_\rmt$ are summarized in Tables~\ref{tab:kappat46}~--~\ref{tab:krt8}.

\input{ktk}

\clearpage
\section{Observables} 
\label{sec:observablefigure}
\newcommand{\evskfiga}[2]{
\begin{figure}[h]
\includegraphics[bb= 0 0 454 340,width=6.5cm]{./#1_plaq.pdf}
\includegraphics[bb= 0 0 454 340,width=6.5cm]{./#1_gact.pdf}\\
\includegraphics[bb= 0 0 454 340,width=6.5cm]{./#1_poly.pdf}
\includegraphics[bb= 0 0 454 340,width=6.5cm]{./#1_chir.pdf}
\caption{\label{fig:#1} #2}
\end{figure}  }

\evskfiga{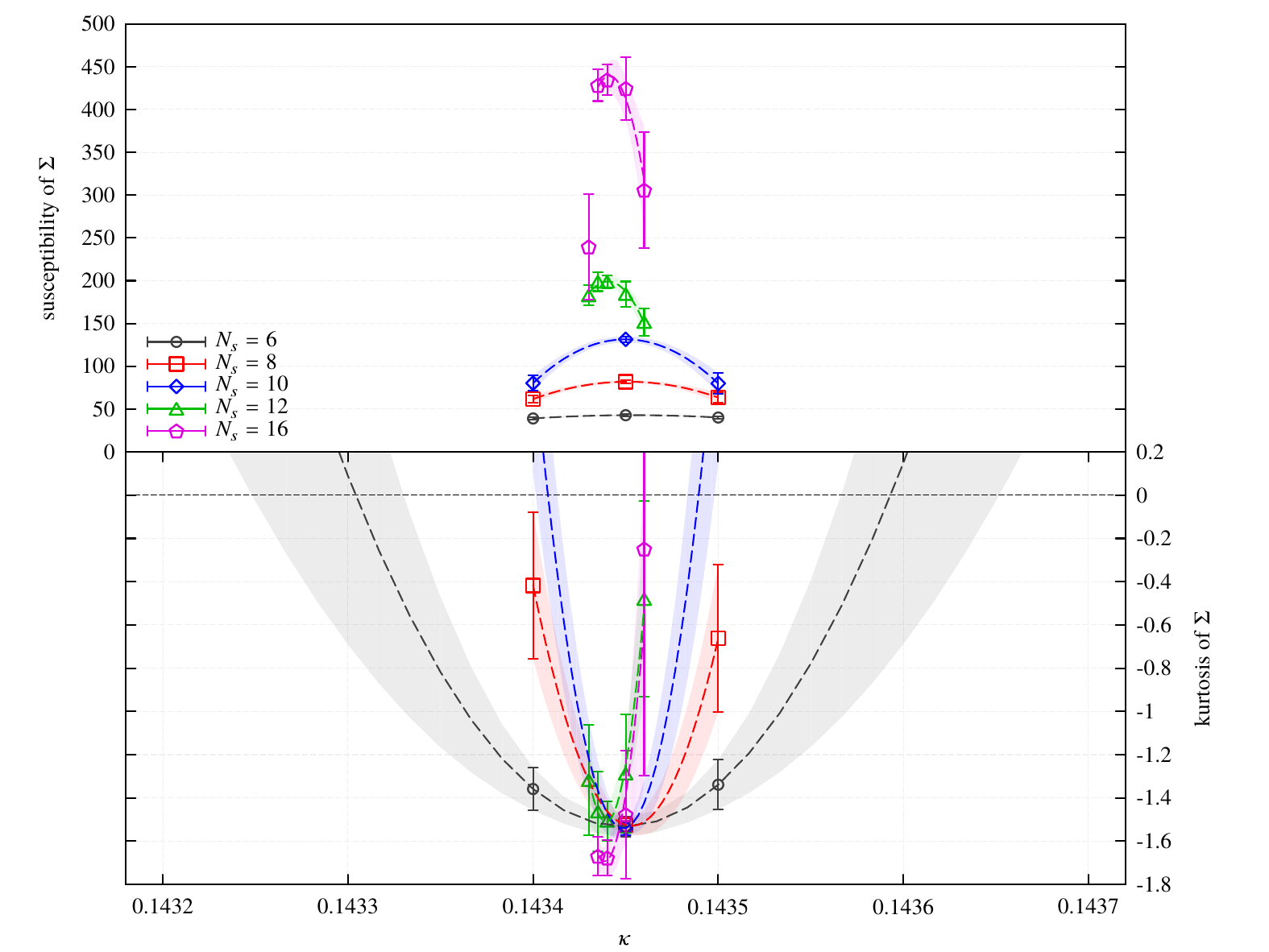}{Susceptibility and kurtosis as functions of $\kappa$ at $\beta=1.60$ and $N_t=4$, together with quadratic fits. Gluon observable $P$ (top left), $G$ (top right), $L$ (bottom right), and quark observable $\Sigma$ (bottom left) are shown. }
\evskfiga{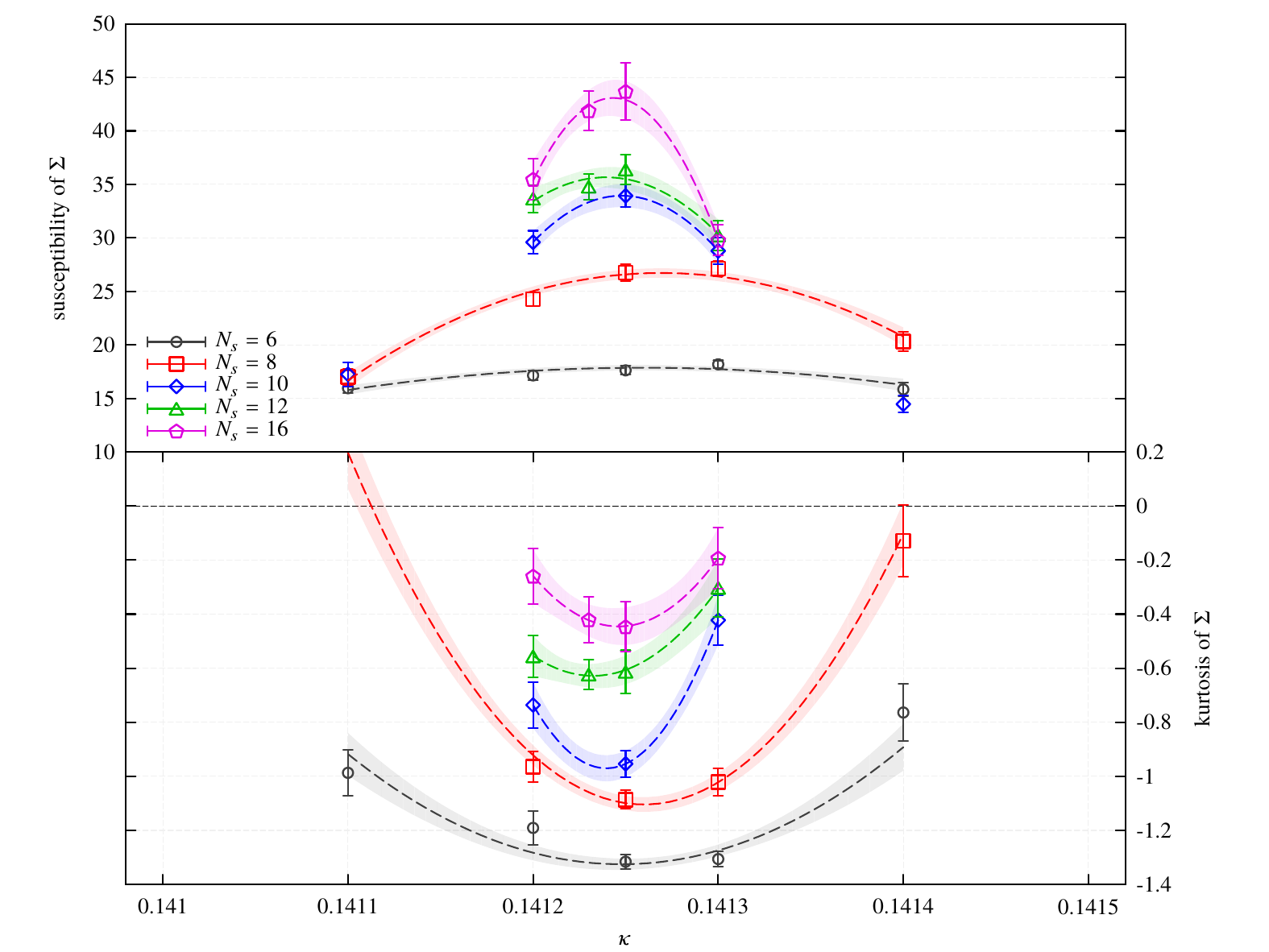}{Same as Fig~\ref{fig:nt4_b160}, but at $\beta=1.65$ and $N_t=4$.}
\evskfiga{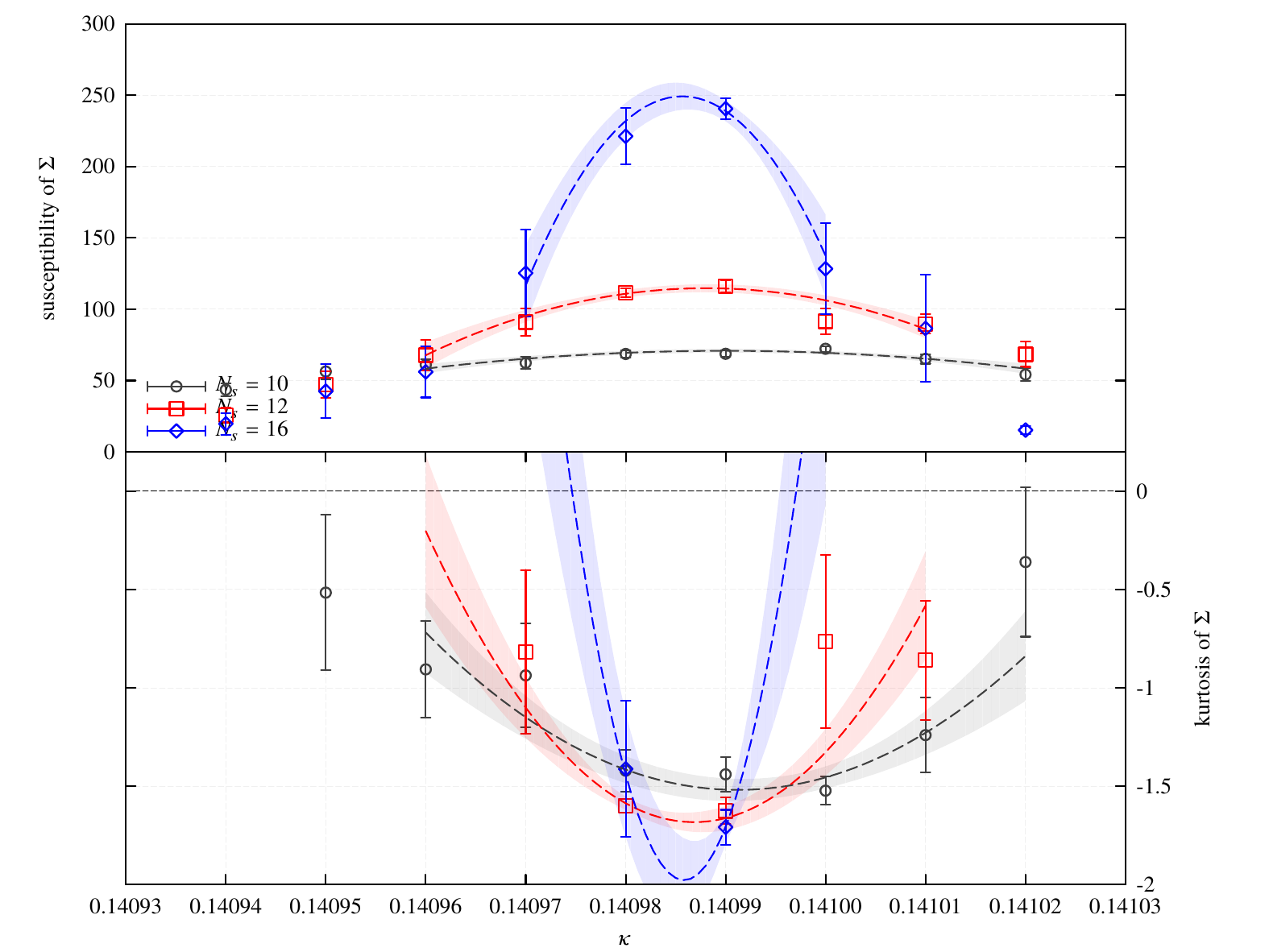}{Susceptibility and kurtosis as functions of $\kappa$ at  $\beta=1.715$ and $N_t=6$, together with quadratic fits.  Gluon observable $P$ (top left), $G$ (top right), $L$ (bottom right), and quark observable $\Sigma$ (bottom left) are shown. }
\evskfiga{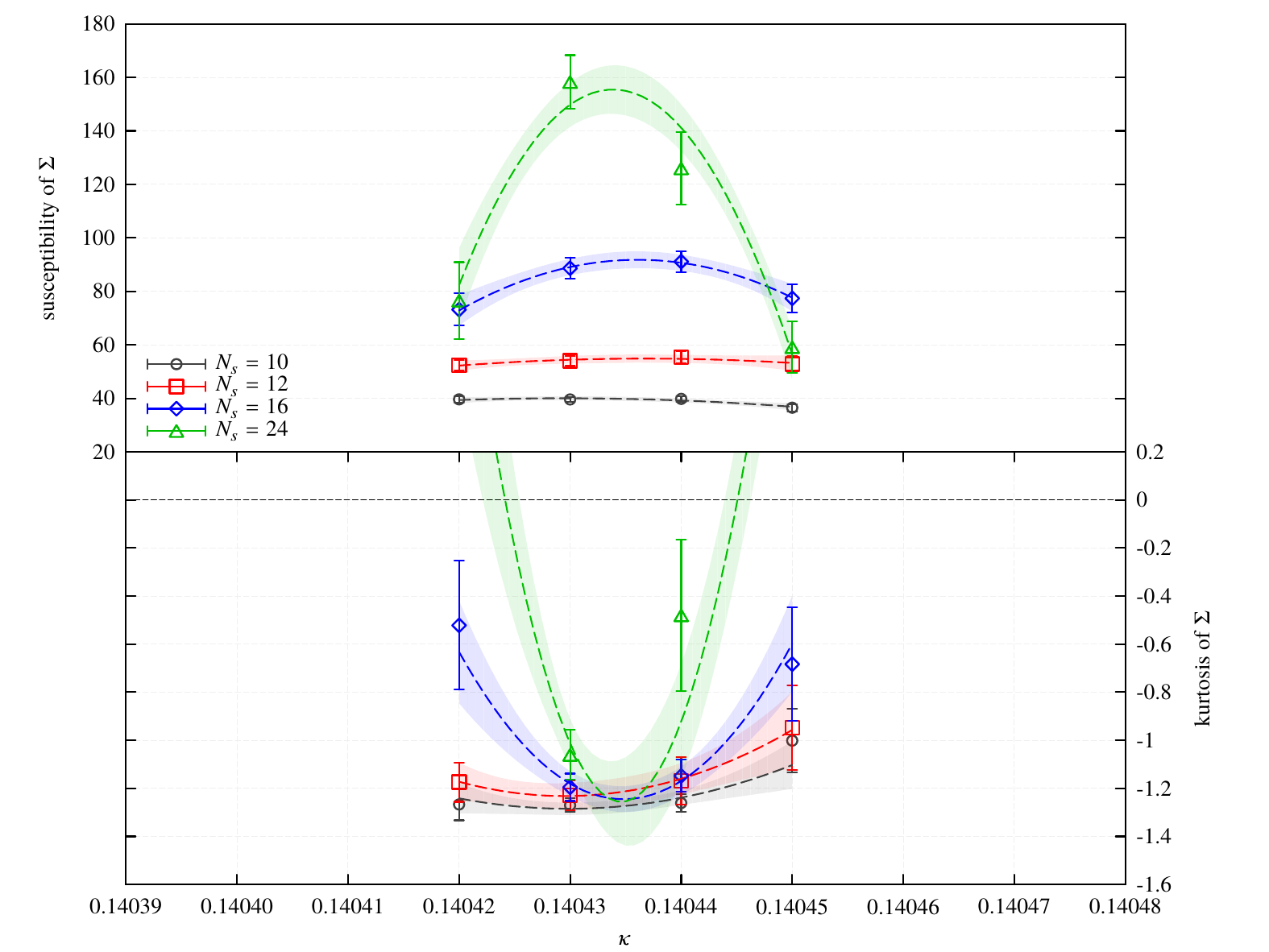}{Same as Fig~\ref{fig:nt6_b1715}, but at $\beta=1.73$ and $N_t=6$.}
\evskfiga{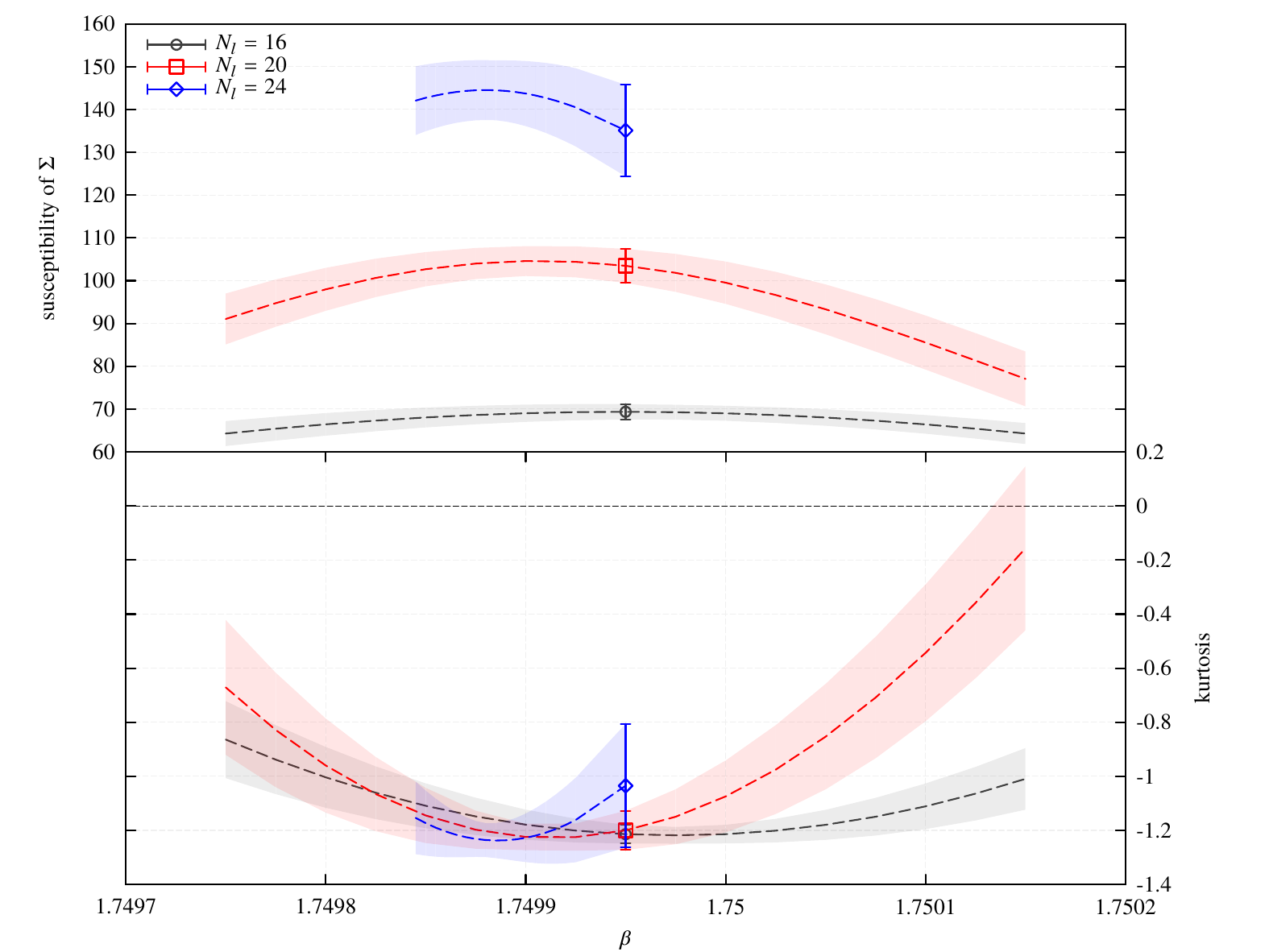}{Susceptibility and  kurtosis as functions of $\beta$ at $\kappa=0.14024$ and $N_t=8$, with curves showing reweighted estimates.  Gluon observable $P$ (top left), $G$ (top right), $L$ (bottom right), and quark observable $\Sigma$ (bottom left) are shown. }
\evskfiga{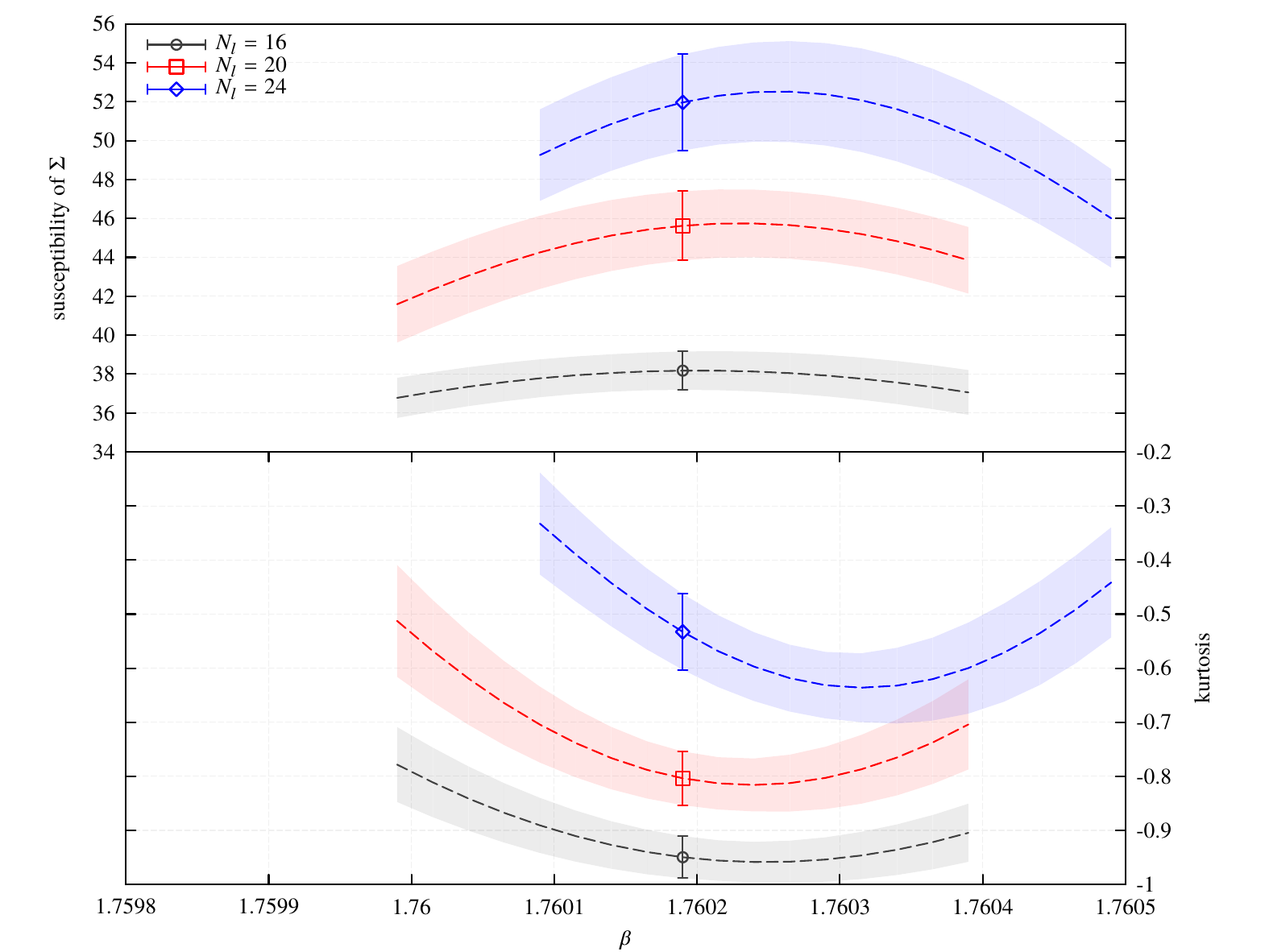}{Same as Fig~\ref{fig:nt8_k014024}, but at $\kappa=0.13995$ and $N_t=8$.}

\end{document}

%% file: scale_table.tex
\begin{table}[h]
\caption{\label{tab:scale160}
Simulation parameters, $\kappa$, $N_l$, $N_t$ and $\sqrt{t_0}/a$ and $am_{\rm PS}$ at $\beta=1.60$.}
\begin{ruledtabular}
\begin{tabular}{ccc|ll}
$\kappa$ & $N_l$ & $N_t$ & $\sqrt{t_0}/a$ &  $am_{\rm PS}$  \\
\colrule
  0.143000 &   12 &   24 &    0.650783(71) &     1.02752(71) \\
  0.143446 &   12 &   24 &    0.653722(72) &     0.98078(68) \\
  0.144000 &   12 &   24 &    0.658485(90) &     0.91122(74) \\
  0.145000 &   12 &   24 &     0.67160(15) &      0.7516(13) \\
\end{tabular}
\end{ruledtabular}
\end{table}

\begin{table}[h]
\caption{\label{tab:scale165}
Simulation parameters, $\kappa$, $N_l$, $N_t$ and $\sqrt{t_0}/a$ and $am_{\rm PS}$ at $\beta=1.65$.}
\begin{ruledtabular}
\begin{tabular}{ccc|ll}
$\kappa$ & $N_l$ & $N_t$ & $\sqrt{t_0}/a$ &  $am_{\rm PS}$  \\
\colrule
  0.140000 &   12 &   24 &    0.659353(65) &     1.17770(72) \\
  0.141240 &   12 &   24 &     0.66818(11) &     1.06023(80) \\
  0.142000 &   12 &   24 &     0.67631(11) &     0.96934(88) \\
  0.143000 &   12 &   24 &     0.69391(21) &      0.8111(14) \\
\end{tabular}
\end{ruledtabular}
\end{table}

\begin{table}[h]
\caption{\label{tab:scale170}
Simulation parameters, $\kappa$, $N_l$, $N_t$ and $\sqrt{t_0}/a$ and $am_{\rm PS}$ at $\beta=1.70$.}
\begin{ruledtabular}
\begin{tabular}{ccc|ll}
$\kappa$ & $N_l$ & $N_t$ & $\sqrt{t_0}/a$ &  $am_{\rm PS}$  \\
\colrule
  0.137100 &   12 &   24 &    0.673734(84) &     1.28517(96) \\
  0.137600 &   12 &   24 &     0.67752(11) &      1.2457(10) \\
  0.138100 &   12 &   24 &     0.68124(11) &      1.1995(11) \\
  0.138250 &   12 &   24 &     0.68277(12) &     1.18585(85) \\
  0.138610 &   12 &   24 &     0.68608(12) &     1.15202(69) \\
  0.140000 &   16 &   32 &    0.705367(99) &     0.99132(64) \\
  0.141000 &   16 &   32 &     0.73207(14) &      0.8243(15) \\
  0.141200 &   16 &   32 &     0.74115(22) &      0.7758(11) \\
  0.141456 &   16 &   32 &     0.75598(23) &      0.7055(12) \\
\end{tabular}
\end{ruledtabular}
\end{table}

\begin{table}[h]
\caption{\label{tab:scale173}
Simulation parameters, $\kappa$, $N_l$, $N_t$ and $\sqrt{t_0}/a$ and $am_{\rm PS}$ at $\beta=1.73$.}
\begin{ruledtabular}
\begin{tabular}{ccc|ll}
$\kappa$ & $N_l$ & $N_t$ & $\sqrt{t_0}/a$ &  $am_{\rm PS}$  \\
\colrule
  0.139000 &   12 &   24 &     0.73453(26) &     0.96412(97) \\
  0.139500 &   12 &   24 &     0.75087(27) &      0.8833(11) \\
  0.140000 &   16 &   32 &     0.77484(31) &      0.7787(11) \\
  0.140334 &   16 &   32 &     0.79915(38) &     0.68974(85) \\
  0.140435 &   16 &   32 &     0.80879(42) &     0.65851(99) \\
  0.140500 &   16 &   32 &     0.81630(36) &     0.63650(93) \\
  0.141000 &   16 &   32 &      0.9391(21) &      0.3306(45) \\
\end{tabular}
\end{ruledtabular}
\end{table}

\begin{table}[h]
\caption{\label{tab:scale175}
Simulation parameters, $\kappa$, $N_l$, $N_t$ and $\sqrt{t_0}/a$ and $am_{\rm PS}$ at $\beta=1.75$.}
\begin{ruledtabular}
\begin{tabular}{ccc|ll}
$\kappa$ & $N_l$ & $N_t$ & $\sqrt{t_0}/a$ &  $am_{\rm PS}$  \\
\colrule
  0.139000 &   12 &   24 &     0.79055(42) &     0.82237(93) \\
  0.139500 &   12 &   24 &     0.82671(76) &      0.7017(15) \\
  0.139529 &   16 &   32 &     0.82959(42) &     0.69470(91) \\
  0.139669 &   16 &   32 &     0.84360(53) &      0.6569(11) \\
  0.139700 &   16 &   32 &     0.84799(45) &     0.64517(96) \\
  0.139850 &   16 &   32 &     0.86861(51) &     0.59293(89) \\
  0.140242 &   16 &   32 &      0.9508(10) &      0.4176(18) \\
\end{tabular}
\end{ruledtabular}
\end{table}

\begin{table}[h]
\caption{\label{tab:scale176}
Simulation parameters, $\kappa$, $N_l$, $N_t$ and $\sqrt{t_0}/a$ and $am_{\rm PS}$ at $\beta=1.76$.}
\begin{ruledtabular}
\begin{tabular}{ccc|ll}
$\kappa$ & $N_l$ & $N_t$ & $\sqrt{t_0}/a$ &  $am_{\rm PS}$  \\
\colrule
  0.139000 &   16 &   32 &     0.83107(33) &     0.73691(77) \\
  0.139500 &   16 &   32 &     0.88650(51) &     0.59667(93) \\
  0.139800 &   16 &   32 &     0.94019(98) &      0.4839(14) \\
  0.139850 &   16 &   32 &      0.9530(12) &      0.4567(17) \\
  0.139950 &   16 &   32 &      0.9823(13) &      0.4060(14) \\
\end{tabular}
\end{ruledtabular}
\end{table}

\begin{table}[h]
\caption{\label{tab:scale177}
Simulation parameters, $\kappa$, $N_l$, $N_t$ and $\sqrt{t_0}/a$ and $am_{\rm PS}$ at $\beta=1.77$.}
\begin{ruledtabular}
\begin{tabular}{ccc|ll}
$\kappa$ & $N_l$ & $N_t$ & $\sqrt{t_0}/a$ &  $am_{\rm PS}$  \\
\colrule
  0.137100 &   12 &   24 &     0.77014(39) &      1.0040(12) \\
  0.137670 &   12 &   24 &     0.79076(35) &     0.91999(86) \\
  0.138500 &   12 &   24 &     0.83773(53) &      0.7675(12) \\
  0.138700 &   12 &   24 &     0.85652(79) &      0.7172(18) \\
  0.138903 &   16 &   32 &     0.87524(52) &     0.66902(80) \\
  0.139000 &   16 &   32 &     0.88795(57) &     0.63966(81) \\
  0.139653 &   16 &   32 &      1.0096(13) &      0.4063(14) \\
  0.139750 &   16 &   32 &      1.0447(17) &      0.3528(20) \\
\end{tabular}
\end{ruledtabular}
\end{table}

\begin{table}[h]
\caption{\label{tab:scalefit} Fit results to  (\ref{eq:scalefit-1}) and (\ref{eq:scalefit-2})  for critical hopping parameter $\kappa_c$ and coefficients for  pseudo scalar meson mass $am_{\rm PS}$ and Wilson flow parameter $\sqrt{t_0}/a$ for $\beta=1.60$ to 1.77. }
\begin{ruledtabular}
\begin{tabular}{c|ccccccc|c}
$\beta$ & $\kappa_c$ & $a_1$ & $a_2$ &$b_0$ &  $b_1$ & $b_2$ & $\chi^2/\dof$  & fit range \\
            &                 &          &         &        &            &         &                  &  $\kappa>$  \\
\colrule
 1.60 &         0.146763(36) &        7.61(17) &       -9.57(74) &      0.7064(13) &      -0.516(14) &       1.149(46)  &   2.18  & 0.1430 \\
 1.65 &         0.145000(29) &        7.60(11) &       -8.01(35) &      0.7409(12) &     -0.5967(99) &       1.079(25)  &  27.55 & 0.1400  \\
 1.70 &         0.142475(26) &       11.16(29) &      -26.1(2.0) &      0.8347(28) &      -1.917(38) &        7.05(18) &      2.32 & 0.1400 \\
 1.73 &        0.1411610(61) &       14.38(18) &      -69.5(2.8) &      0.9804(35) &       -6.65(14) &       53.8(1.4)   &  31.94 & 0.1400  \\
 1.75 &        0.1405946(61) &       10.18(12) &      -23.3(1.3) &      1.0346(26) &      -5.465(70) &       30.39(58) &  12.92 & 0.1390   \\
 1.76 &        0.1403027(71) &        9.61(18) &      -22.2(2.3) &      1.0791(34) &       -6.29(13) &       38.7(1.3)   &  6.75   & 0.1390 \\
 1.77 &        0.1400560(95) &        8.21(25) &      -11.7(3.4) &      1.1315(41) &       -6.47(14) &       36.2(1.5)   &  7.74   & 0.1387  \\
\end{tabular}
\end{ruledtabular}
\end{table}

%% file: ktk.tex

\begin{table}[h]
\caption{\label{tab:kappat46}
Transition point for each observable at $N_t=4, 6$.
}
\begin{ruledtabular}
\begin{tabular}{rr|lllll}
$N_t$ & $N_l$ & $\beta$ & $\kappa_{\rmt,P}$ & $\kappa_{\rmt,G}$ & $\kappa_{\rmt,L}$ & $\kappa_{\rmt,\Sigma}$ \\
\colrule
 4 &        6 & 1.60 & 0.1434492(75)   & 0.1434487(75)   & 0.1434445(86)   & 0.1434543(77)   \\
 4 &        8 & 1.60 & 0.1434481(48)   & 0.1434478(48)   & 0.1434466(48)   & 0.1434512(48)   \\
 4 &       10 & 1.60 & 0.1434485(36)   & 0.1434484(36)   & 0.1434473(36)   & 0.1434499(37)   \\
 4 &       12 & 1.60 & 0.1434386(36)   & 0.1434385(36)   & 0.1434379(36)   & 0.1434404(30)   \\
 4 &       16 & 1.60 & 0.1434392(41)   & 0.1434392(41)   & 0.1434387(42)   & 0.1434412(40)   \\
\colrule
 4 &        6 & 1.65 & 0.1412184(94)   & 0.1412158(96)   & 0.1412102(96)   & 0.141260(16)    \\
 4 &        8 & 1.65 & 0.1412541(56)   & 0.1412529(56)   & 0.1412450(53)   & 0.1412695(58)   \\
 4 &       10 & 1.65 & 0.1412430(45)   & 0.1412426(46)   & 0.1412381(54)   & 0.1412479(43)   \\
 4 &       12 & 1.65 & 0.1412321(67)   & 0.1412314(69)   & 0.1412266(76)   & 0.1412394(80)   \\
 4 &       16 & 1.65 & 0.1412416(25)   & 0.1412413(26)   & 0.1412379(30)   & 0.1412433(36)   \\
\colrule
 4 &        6 & 1.70 & 0.138633(11)    & 0.138627(11)    & 0.1386224(94)   & \\
 4 &        8 & 1.70 & 0.138636(21)    & 0.138631(20)    & 0.138615(18)    & \\
 4 &       10 & 1.70 & 0.1386188(99)   & 0.1386157(98)   & 0.1386081(84)   & \\
\colrule
 6 &       10 & 1.715 & 0.1409862(28)   & 0.1409861(28)   & 0.1409945(26)   & 0.1409900(26)   \\
 6 &       12 & 1.715 & 0.1409867(16)   & 0.1409867(16)   & 0.1409893(15)   & 0.1409881(17)   \\
 6 &       16 & 1.715 & 0.1409853(12)   & 0.1409853(12)   & 0.1409860(12)   & 0.1409856(12)   \\
\colrule
 6 &       10 & 1.73 & 0.1404250(90)   & 0.1404248(93)   & 0.1404360(38)   & 0.1404295(55)   \\
 6 &       12 & 1.73 & 0.1404333(46)   & 0.1404333(45)   & 0.1404389(64)   & 0.1404367(69)   \\
 6 &       16 & 1.73 & 0.1404347(17)   & 0.1404346(17)   & 0.1404365(17)   & 0.1404361(17)   \\
 6 &       24 & 1.73 & 0.14043436(53)  & 0.14043435(53)  & 0.14043496(52)  & 0.14043387(80)  \\
\colrule
 6 &       10 & 1.75 & 0.1396579(92)   & 0.1396575(94)   & 0.1396804(44)   & 0.1396564(85)   \\
 6 &       12 & 1.75 & 0.1396607(72)   & 0.1396605(74)   & 0.1396777(68)   & 0.1396673(69)   \\
 6 &       16 & 1.75 & 0.1396588(81)   & 0.1396584(82)   & 0.1396716(42)   & 0.1396676(41)   \\
\colrule
 6 &       10 & 1.77 & 0.1388967(85)   & 0.1388958(87)   & 0.1389364(98)   & \\
 6 &       12 & 1.77 & 0.1388934(49)   & 0.1388928(49)   & 0.1389263(61)   & \\
 6 &       16 & 1.77 & 0.1388942(46)   & 0.1388937(47)   & 0.1389270(59)   & \\
\end{tabular}
\end{ruledtabular}
\end{table}

\begin{table}[h]
\caption{\label{tab:betat8}Transition point for each observable at $N_t=8$.}
\begin{ruledtabular}
\begin{tabular}{rr|lllll}
$N_t$ & $N_l$ & $\kappa$ & $\beta_{\rmt,P}$ & $\beta_{\rmt,G}$ & $\beta_{\rmt,L}$ & $\beta_{\rmt,\Sigma}$ \\
\colrule
 8 &       12 & 0.14054 & 1.739825(75)    & 1.739825(75)    & 1.740175(99)    & \\
 8 &       16 & 0.14054 & 1.739930(46)    & 1.739930(46)    & 1.739972(46)    & \\
\colrule
 8 &       12 & 0.14024 & 1.749791(96)    & 1.749791(96)    & 1.75033(12)     & \\
 8 &       16 & 0.14024 & 1.749894(28)    & 1.749894(28)    & 1.749997(27)    & 1.749950(19)    \\
 8 &       20 & 0.14024 & 1.749906(22)    & 1.749906(22)    & 1.749940(22)    & 1.749909(14)    \\
 8 &       24 & 0.14024 & 1.749887(19)    & 1.749887(19)    & 1.749905(19)    & 1.749880(22)    \\
\colrule
 8 &       12 & 0.13995 & 1.75975(11)     & 1.75975(11)     & 1.76047(13)     & \\
 8 &       16 & 0.13995 & 1.760098(31)    & 1.760098(31)    & 1.760305(32)    & 1.760201(35)    \\
 8 &       20 & 0.13995 & 1.760185(26)    & 1.760185(26)    & 1.760268(27)    & 1.760230(30)    \\
 8 &       24 & 0.13995 & 1.760207(35)    & 1.760207(35)    & 1.760294(35)    & 1.760255(25)    \\
\colrule
 8 &       12 & 0.13966 & 1.769664(99)    & 1.769663(99)    & 1.77086(14)     & \\
 8 &       16 & 0.13966 & 1.769956(88)    & 1.769955(88)    & 1.770396(96)    & \\
 8 &       20 & 0.13966 & 1.770061(65)    & 1.770062(65)    & 1.770267(74)    & \\
\colrule
 8 &       12 & 0.1394 & 1.77816(15)     & 1.77816(15)     & 1.77995(16)     & \\
 8 &       16 & 0.1394 & 1.778857(96)    & 1.778857(96)    & 1.77956(13)     & \\
 8 &       20 & 0.1394 & 1.77893(11)     & 1.77892(12)     & 1.77933(14)     & \\
\end{tabular}
\end{ruledtabular}
\end{table}

\begin{table}[h]
\caption{\label{tab:krt46}
Value of kurtosis at transition point for each observable at $N_t=4, 6$.
}
\begin{ruledtabular}
\begin{tabular}{rr|lllll}
$N_t$ & $N_l$ & $\beta$ & $K_{\rmt,P}$ & $K_{\rmt,G}$ & $K_{\rmt,L}$ & $K_{\rmt,\Sigma}$ \\
\colrule
 4 &        6 & 1.60 & -1.231(45)      & -1.219(46)      & -1.103(39)      & -1.530(47)      \\
 4 &        8 & 1.60 & -1.401(29)      & -1.396(28)      & -1.303(26)      & -1.528(40)      \\
 4 &       10 & 1.60 & -1.451(46)      & -1.449(48)      & -1.386(53)      & -1.540(36)      \\
 4 &       12 & 1.60 & -1.466(56)      & -1.466(55)      & -1.434(47)      & -1.508(84)      \\
 4 &       16 & 1.60 & -1.628(58)      & -1.628(57)      & -1.609(52)      & -1.677(87)      \\
\colrule
 4 &        6 & 1.65 & -0.911(14)      & -0.899(14)      & -0.877(13)      & -1.323(22)      \\
 4 &        8 & 1.65 & -0.833(26)      & -0.823(26)      & -0.792(22)      & -1.100(28)      \\
 4 &       10 & 1.65 & -0.792(45)      & -0.788(45)      & -0.755(41)      & -0.960(50)      \\
 4 &       12 & 1.65 & -0.583(26)      & -0.579(25)      & -0.569(24)      & -0.625(46)      \\
 4 &       16 & 1.65 & -0.380(41)      & -0.377(41)      & -0.354(38)      & -0.445(69)      \\
\colrule
 4 &        6 & 1.70 & -0.617(16)      & -0.600(16)      & -0.677(15)      & \\
 4 &        8 & 1.70 & -0.441(29)      & -0.432(28)      & -0.479(27)      & \\
 4 &       10 & 1.70 & -0.333(35)      & -0.327(34)      & -0.296(32)      & \\
\colrule
 6 &       10 & 1.715 & -1.308(56)      & -1.299(56)      & -1.243(32)      & -1.518(57)      \\
 6 &       12 & 1.715 & -1.529(32)      & -1.523(32)      & -1.446(79)      & -1.680(51)      \\
 6 &       16 & 1.715 & -1.90(21)       & -1.89(21)       & -1.79(17)       & -1.98(22)       \\
\colrule
 6 &       10 & 1.73 & -1.044(31)      & -1.034(31)      & -1.004(37)      & -1.285(26)      \\
 6 &       12 & 1.73 & -1.034(53)      & -1.026(53)      & -0.997(56)      & -1.198(74)      \\
 6 &       16 & 1.73 & -1.122(52)      & -1.117(52)      & -1.108(50)      & -1.240(55)      \\
 6 &       24 & 1.73 & -1.14(14)       & -1.14(14)       & -1.08(14)       & -1.25(18)       \\
\colrule
 6 &       10 & 1.75 & -0.638(17)      & -0.625(17)      & -0.737(21)      & -1.008(39)      \\
 6 &       12 & 1.75 & -0.616(24)      & -0.605(24)      & -0.698(26)      & -0.879(37)      \\
 6 &       16 & 1.75 & -0.431(33)      & -0.425(33)      & -0.479(33)      & -0.618(50)      \\
\colrule
 6 &       10 & 1.77 & -0.353(31)      & -0.340(30)      & -0.639(36)      & \\
 6 &       12 & 1.77 & -0.316(38)      & -0.308(37)      & -0.457(40)      & \\
 6 &       16 & 1.77 & -0.226(31)      & -0.220(31)      & -0.233(42)      & \\
\end{tabular}
\end{ruledtabular}
\end{table}

\begin{table}[h]
\caption{\label{tab:krt8}
Value of kurtosis at transition point for each observable at $N_t=8$.
}
\begin{ruledtabular}
\begin{tabular}{rr|lllll}
$N_t$ & $N_l$ & $\kappa$ & $K_{\rmt,P}$ & $K_{\rmt,G}$ & $K_{\rmt,L}$ & $K_{\rmt,\Sigma}$ \\
\colrule
 8 &       12 & 0.14054 & -1.126(32)      & -1.118(32)      & -0.41(60)       & \\
 8 &       16 & 0.14054 & -1.411(23)      & -1.406(23)      & -1.17(17)       & \\
\colrule
 8 &       12 & 0.14024 & -0.856(37)      & -0.847(37)      & -0.29(48)       & \\
 8 &       16 & 0.14024 & -1.010(25)      & -1.003(25)      & -0.93(12)       & -1.214(17)      \\
 8 &       20 & 0.14024 & -1.103(32)      & -1.098(32)      & -0.923(80)      & -1.226(28)      \\
 8 &       24 & 0.14024 & -1.126(51)      & -1.122(51)      & -1.03(10)       & -1.236(29)      \\
\colrule
 8 &       12 & 0.13995 & -0.581(38)      & -0.570(38)      & -0.20(45)       & \\
 8 &       16 & 0.13995 & -0.684(24)      & -0.676(24)      & -0.74(16)       & -0.953(16)      \\
 8 &       20 & 0.13995 & -0.648(34)      & -0.642(34)      & -0.65(17)       & -0.815(20)      \\
 8 &       24 & 0.13995 & -0.488(64)      & -0.483(64)      & -0.43(21)       & -0.611(29)      \\
\colrule
 8 &       12 & 0.13966 & -0.399(37)      & -0.388(37)      & -0.57(53)       & \\
 8 &       16 & 0.13966 & -0.389(53)      & -0.381(52)      & -0.48(41)       & \\
 8 &       20 & 0.13966 & -0.357(57)      & -0.351(56)      & -0.21(32)       & \\
\colrule
 8 &       12 & 0.1394 & -0.247(50)      & -0.238(50)      & -0.26(30)       & \\
 8 &       16 & 0.1394 & -0.240(37)      & -0.234(36)      & -0.79(30)       & \\
 8 &       20 & 0.1394 & -0.214(47)      & -0.206(47)      & -0.08(30)       & \\
\end{tabular}
\end{ruledtabular}
\end{table}